\documentclass[fleqn,
12pt,
twoside]{amsart}

\usepackage{amssymb,amsfonts,latexsym}

\begin{document}
  
\baselineskip=20pt

\def\op{\sharp}
\def\pop{Q}
\def\Al{\Omega_0}

\def\A{{\mathcal A}}
\def\Aa{{\mathfrak A}}
\def\AA{A^\sharp}
\def\alphc{\rho_{W}}

\def\B{{  B}}
\def\C{{\Bbb C}}
\def\Cb{{  C}}
\def\delq{\phi}
\def\delp{\chi}
\def\H{{  H}}
\def\I{{\mathfrak I}}
\def\IPa{}
\def\IQa{  }
\def\IP{ }
\def\IQ{ }
\def\IbPa{ }
\def\IbQa{ }
\def\IbP{ }
\def\IbQ{ }
\def\mcJ{{\mathcal J}}
\def\Jj{\mcJ }
\def\K{A_a}
\def\Mm{{\mathcal M}}
\def\N{{\bf N}}
\def\R{{\Bbb R}}
\def\rV{{{\rm deg}(V)}}
\def\S{{\Bbb S}}
\def\Z{{\Bbb Z}}
\def\Zb{{ Z}}
\def\T{{\mathcal T}}
\def\TT{{\Bbb T}}
\def\Q{{\mathcal Q}}
\def\xa{a}

\def\alph{\rho_W}
\def\bbeta{\bar{\rho}_W}

\def\PV{\pi_V}
\def\PVc{\bar{\pi}_V}
\def\sig{s}
\def\tPhi{\tilde{\Phi}}
\def\cL{{\mathcal L}}
\def\cF{{\Lambda}}
\def\tJ{\tilde{J}}

\def\1{{\bf 1}}
\def\phys{{\mathcal  S}}
\def\grad{\nabla_g}
\def\crit{{\mathfrak C}}
\def\cg{\crit_{gen}}
\def\ker{{\rm ker}}

\def\eqnn{\begin{eqnarray*}}
\def\eeqnn{\end{eqnarray*}}
\def\eqn{\begin{eqnarray}}
\def\eeqn{\end{eqnarray}}
\def\nbf{}
\def\nit{}

\def\bc{\begin{center}}
\def\ec{\end{center}}

\newtheorem{thm}{Theorem}[section]
\newtheorem{defn}{Definition}[section]
\newtheorem{prop}{Proposition}[section]
\newtheorem{hyp}{Hypothesis}[section]
\newtheorem{lem}{Lemma}[section]
\newtheorem{cor}{Corollary}[section]
\newtheorem{conj}{Conjecture}[section]

\title{  CRITICAL MANIFOLDS
AND  
STABILITY IN HAMILTONIAN
SYSTEMS  WITH NON-HOLONOMIC CONSTRAINTS }
\author{Thomas Chen} 
\address{
Courant Institute of Mathematical Sciences, New York University,
251 Mercer Street,
New York, NY 10012-1185.
chenthom@cims.nyu.edu }
\date{}

\maketitle
 
\begin{abstract} 
We explore a particular approach to the analysis of dynamical and
geometrical properties of autonomous, Pfaffian 
non-holonomic systems in classical mechanics. 
The method is based on the construction of  
a certain auxiliary constrained Hamiltonian system, which comprises the
non-holonomic mechanical system as a dynamical subsystem on an invariant manifold. 
The embedding system possesses a completely natural structure in the context 
of symplectic geometry, and using it in order to understand
properties of the subsystem has compelling advantages.  
We discuss generic geometric and topological properties of the critical sets of both 
embedding and physical system, using Conley-Zehnder theory and by relating
the Morse-Witten complexes of the 'free' and constrained system to one another.
Furthermore, we give a qualitative discussion of the stability of motion in
the vicinity of the critical set. We point out key relations
to sub-Riemannian geometry, and a potential computational application.
\end{abstract}

\section*{Introduction}

We introduce and explore 
a particular approach to the analysis of autonomous, Pfaffian 
non-holonomic systems in classical mechanics, which renders them naturally
accessible to the methods of symplectic and sub-Riemannian geometry. 
We note that typical examples of systems encountered in sub-Riemannian geometry 
emerge from optimal control, or 'vakonomic' problems, which are derived from
a different variational principle (minimization of the Carnot-Caratheodory
distance) than the Euler-Lagrange equations of classical mechanical systems with 
non-holonomic constraints (the H\"older variational principle, cf. \cite{Ar1}, 
and section III in this paper).
The strategy is based on the introduction of an artificial Hamiltonian system
with constraints that are compatible with the symplectic structure,
constructed in a manner that it comprises the non-holonomic mechanical system 
as a dynamical
subsystem on an invariant manifold.  
The main focus of the discussion in this paper aims at the geometrical and topological
properties of the critical sets of 
both  embedding and mechanical system, on the stability of equilibria, and  
an application of the given analysis to a computational problem.

There exists a multitude of different approaches to
the description and analysis of
non-holonomic systems in classical mechanics, stemming from
various subareas of application.
The geometrical approach given here has been
strongly influenced by \cite{We} and   
\cite{Bra,Bra2,SoBr}.
A construction for the Lagrangian case, which is closely related to
what will be presented in section III,  has been
given in \cite{CaFa}. A different approach
in the Hamiltonian picture is dealt with in \cite{vdSMa}.
A geometrical theory of non-holonomic systems with a strong
influence of network theory has been developed in \cite{YoKa}.
The geometrical structure of non-holonomic systems
with symmetries and the associated reduction theory, as well as
aspects of their stability theory has been at the focus in
the important works   \cite{BlKrMaMu,KoMa,KoMa1,ZeBlMa}, and other 
papers by the same authors.

This paper is structured as follows.
In section I, we introduce a  class of 
Hamiltonian systems with non-integrable constraints.
Given a symplectic manifold $(M,\omega)$
and
a non-integrable,
symplectic distribution $V$, we focus on the flow $\tPhi_t$
generated by the component $X_H^V$
of the Hamiltonian vector field $X_H$ in $V$.
In section II, we study the geometry and topology of the
critical set $\crit$ of the constrained
Hamiltonian system.
The main technical tool used for this purpose is a gradient-like flow
$\phi_t$,
whose critical set $\crit$ is identical to that of $\tPhi_t$.
Assuming that the Hamiltonian $H:M\rightarrow\R$ is a Morse
function, it is proved that generically, $\crit$ is a normal hyperbolic
submanifold of $M$.
Using Conley-Zehnder theory, we prove a topological formula for closed, compact $\crit$,
that is closely related to the Morse-Bott inequalities.
A second, alternative proof is given, based on the use of the Morse-Witten
complex, to elucidate relations between
the 'free' and the constrained system.
In section III, we give a qualitative, partly non-rigorous discussion of the stability 
of the constrained Hamiltonian system, 
and conjecture a
stability criterion for the critically stable case. A proof of the asserted criterion, which would involve methods of KAM and Nekhoroshev theory, is beyond the scope of the present work.  
We derive an expression for 
orbits in the vicinity of a critically stable equilibrium that is
adapted to the flag of $V$, and 
point out relations  
to sub-Riemannian geometry.

In section IV, we consider
Hamiltonian mechanical systems with Pfaffian constraints.
We show that for any such system, there exists an
auxiliary constrained Hamiltonian system of the type introduced
in section I. We study the global topology of
the critical manifold of the constrained mechanical system, and
again discuss
the stability of equilibria.
Finally,
we propose a computational application, a method to
numerically determine the generic connectivity components
of the critical manifold.


\section{A NON-INTEGRABLE GENERALIZATION OF  
DIRAC CONSTRAINTS}
\label{gendir}

Let $(M,\omega,H)$ be a Hamiltonian system, where $M$ is a smooth,
symplectic $2n$-manifold with $C^\infty$ symplectic structure
$\omega\in \Lambda^2(M)$, and where
$H\in C^\infty(M)$ is the Hamiltonian function. For $p=1,\dots,2n$,
let $\Lambda^p(M)$ denote the $C^\infty(M)$-module of $p$-forms on $M$.
The Hamiltonian vector field $X_H\in \Gamma(TM)$ is determined by
$$i_{X_H}\omega=-dH\;,$$
where $i$ stands for interior multiplication.
Given a smooth distribution $W\subset TM$,
$\Gamma(W)$ will denote the $C^\infty(M)$-module
of smooth sections of $W$.
The Hamiltonian flow is the 1-parameter group
$\Phi_t\in {\rm Diff}(M)$ generated by $X_{H}$, with $t\in \R$,
and $\Phi_0={\rm id}$.
Its orbits are solutions of
\eqn\label{hameqmo}\partial_t\Phi_t(x)\;=\;X_H(\Phi_t(x))\;\eeqn
for all $x\in M$, and $t\in\R$.

Let us first recall some standard
facts about Dirac constraints that will be useful in the subsequent
construction. 
Let, for $f,g\in C^\infty$,
\eqn\label{poisson}\lbrace f,g\rbrace=\omega(X_f,X_g)\; \eeqn
denote the smooth, non-degenerate Poisson structure on $M$ induced 
by $\omega$.
It is a derivative in both of its arguments,
and satisfies the Jacobi identity
$\lbrace f,\lbrace g,h\rbrace\rbrace+($cyclic$) = 0$, 
thus $(C^\infty(M),\lbrace\cdot,\cdot\rbrace)$ is a Lie algebra.
Then, (~\ref{hameqmo}) translates into
\eqn\label{Poisseqsmo}\partial_t f(\Phi_t(x))\;=\;
     \lbrace H, f\rbrace(\Phi_t(x)) \; , \eeqn
for all $f\in C^\infty(M)$, and all $x\in M$, $t\in\R$.

Let $j:M'\hookrightarrow M$ be an embedded, smooth,
$2k$-dimensional symplectic submanifold of $M$,
endowed with the pullback symplectic
structure $j^*\omega$. The Dirac bracket corresponds to the induced
Poisson bracket on $M'$,  
$$\lbrace f,g\rbrace_D\;=\;
      (j^*\omega)(X_{\tilde{f}},X_{\tilde{g}}) \; ,$$
defined for any pair of extensions $\tilde{f},\tilde{g}\in C^\infty(M)$ of
$f,g\in C^\infty(M')$.
If $M'$ is locally characterized as the locus of common
zeros of some family of functions $G_i\in C^\infty(M)$,  
$i=1,\dots,2(n-k)$,   the following
explicit construction of the Dirac bracket can be given,  \cite{MaRa}.
Since $M'\subset M$ is symplectic, the $(n-k)^2$ functions locally
given by
$D_{ij}:=\lbrace G_i,G_j\rbrace$ can be patched together to
define a matrix-valued $C^\infty$ function 
that
is invertible everywhere on $M'$.
The explicit formula for the Dirac bracket is locally
given by
\eqn\label{dirbra}\lbrace f,g\rbrace_D\;=\;
       \lbrace \tilde{f},\tilde{g}\rbrace\;-
      \;\lbrace \tilde{f},G_i\rbrace
      \;D^{ij}\;\lbrace G_j,\tilde{g}\rbrace\;,\eeqn
where $D^{ij}$ denotes the components of the inverse of
$[D_{ij}]$.

This construction can be put into the more general context 
of a symplectic distribution.

\begin{defn}
A distribution $V$ over the base manifold $M$ is symplectic
if $V_x$ is a symplectic subspace of $T_x M$ with respect to 
$\omega_x$, for all $x\in M$. Its symplectic complement
$V^\perp$ is the distribution which is fibrewise
$\omega$-skew orthogonal to $V$.  
Furthermore, an embedding $I\subset\R\hookrightarrow M$ that is tangent 
to $V$ is called $V$-horizontal. 
\end{defn}

Clearly, $V^\perp$ is by itself symplectic, and
smoothness of $V$ and $\omega$ implies smoothness
of  $V^\perp$. Furthermore, the Whitney sum bundle
$V\oplus V^\perp$ is $TM$.
Thus, let $V$ denote an
integrable, smooth, symplectic rank $2k$-distribution $V$ over $M$.
Clearly, any section $X\in\Gamma(TM)$ has a decomposition
$$X=X^V + X^{V^\perp}\;,$$
where $X^{V^{(\perp)}}\in\Gamma(V^{(\perp)})$, so that
$\omega(X^V , X^{V^\perp})=0$.
Furthermore, there exists an $\omega$-skew orthogonal tensor
$\PV:TM\longrightarrow TM$ with
$${\rm Ker}(\PV)=V^\perp\;\;\;\;,
    \;\;\;\;\PV(X)=X\;\;\;\;\forall X\in\Gamma(V)\;,$$
which
satisfies
\eqn\omega(\PV (X),Y)=\omega(X,\PV (Y))\label{omskewsymmeq}\eeqn
for all $X,Y\in\Gamma(TM)$.
It will be referred to as the $\omega$-skew orthogonal
projection tensor associated to $V$.
Let
$Y_1,\dots,Y_{2k}$ denote a local spanning family of vector fields
for $V$. Then,   $V$ being symplectic is
equivalent to  the matrix $[C_{ij}]:=[\omega(Y_i,Y_j)]$ being invertible.

\begin{lem}\label{Pconstr}
Let $[C^{kl}]$ denote the inverse of $[C_{ij}]$, and let
$\theta_j :=i_{Y_j}\omega \in\Lambda^1(TM)$.
Then, locally,
$\PV=C^{ij}Y_i\otimes\theta_j$.
\end{lem}

\begin{proof}
The fact that $C^{ij}Y_i\otimes\theta_j$ is a projector, and that
(~\ref{omskewsymmeq}) holds,
follows from $\theta_i(Y_j)=C_{ij}$, and 
$C_{ij}C^{jk}=\delta_i^l$. Its rank is clearly $2k$, and it is
straightforward to see that its kernel is given by 
$\Gamma(V^\perp)$. \end{proof}

\subsection{\bf Non-Integrable Constraints}

The quadruple $(M,\omega,H,V)$ naturally defines a  
dynamical system whose orbits are all $V$-horizontal.  Its flow is simply
the 1-parameter group of diffeomorphisms  $\tPhi_t$
generated by $X_H^V:=\PV( X_H)\in\Gamma(V)$, with
\eqn
	\partial_t \tPhi_t(x) = X_H^V (\tPhi_t(x) )  
\label{eqsofmo}
\eeqn
for every $x\in M$.
In a local Darboux chart, where $\omega$ is
represented by
\eqn\label{sympJ}\mcJ \;=\;
      \left(\begin{array}{cc}0&\1_n\\-\1_n&0\end{array}\right)\;,\eeqn
and where  $x(t)$ stands for the vector of coordinate components of 
$\tPhi_t(\xa)$,  
(~\ref{eqsofmo}) is given by
\eqn\partial_t x(t) =
     \left(P_x\;\mcJ \;\partial_x H\right)(x(t)) 
    = \left(\mcJ \; P_x^\dagger\;\partial_x H\right)(x(t))\;.
    \label{eqsmocoord}\eeqn
$P$ denotes the matrix of $\PV$, and $P^\dagger$ is
its transpose.

If the condition of integrability
imposed on $V$ is dropped,
this dynamical system will allow for the description of
non-holonomic mechanics.
If $V$ is integrable, $M$ is foliated into
$2k$-dimensional symplectic integral manifolds of $V$.
On every leaf $j:M'\rightarrow M$, the induced dynamical system 
corresponds to the
pullback Hamiltonian system $(M,j^*\omega,H\circ j)$.
In this sense, (~\ref{eqsofmo}) generalizes the
Dirac constraints.

A new class of dynamical systems is obtained by discarding the
requirement of integrability on  $V$.
Let $[\cdot,\cdot]$ denote the Lie bracket.
We recall that the distribution $V$ is non-integrable if there
exists a filtration
\eqn
	\label{filt}V_0\subset V_1\subset V_2\subset\cdots\subset V_r,
\eeqn
inductively defined by $V_0=V$, and
$V_i=V_{i-1}+[V_0,V_{i-1}]$, where $V_1\neq V_0$.
The sequence $\lbrace V_i\rbrace_1^r$ is called the {\em flag} of $V$. 
If the fibre ranks of all $V_i$ are base point independent,
$V$ is called equiregular.
The smallest number $\rV $ at which the flag of $V$
stabilizes, that is, for which
$V_s=V_{\rV }$, $\forall s\geq \rV $,
is called the
degree of non-holonomy of $V$.
If $V_{\rV }=TM$, one says that $V$ satisfies Chow's condition, or 
that it is totally non-holonomic.

\begin{prop}\label{Frob}(Frobenius condition)
$V$ is integrable if and only if locally,
\eqn\cF^k_{ij} :=  (\PV)^r_i (\PV)^s_j \left(\partial_r(\PV)^k_s-
      \partial_s(\PV)^k_r\right) = 0 \eeqn
everywhere on $M$.
\end{prop}

\begin{proof}
$V$ is integrable if and only if
$\PVc\left([\PV (X),\PV (Y)]\right)=0$
for all sections $X,Y$ of $TM$, which is equivalent
to $V_1=V$. The asserted
formula is the local coordinate representation of this
condition. \end{proof}

\subsection{An  Auxiliary Almost K\"ahler Structure }

For the analysis in subsequent sections, it will be useful to 
introduce an almost K\"ahler structure on $M$ that is adapted
to $V$. To this end, let us briefly recall some basic definitions.
Let $g$ denote a Riemannian metric
on $M$.
An almost complex structure $J$ is a smooth bundle
isomorphism $J:TM\rightarrow TM$ with
$J^2=-\1$. Together with $g$, it defines a two form satisfying
\eqn\label{sksadness}\omega_{g,J}(X,Y)=g(JX,Y) \eeqn
for all sections $X$, $Y\in\Gamma(TM)$.
$g$ is hermitean if
$g(JX,JY)=g(X,Y)$, and K\"ahler if $\omega_{g,J}$
is closed. The triple $(g,J,\omega_{g,J})$  
is called compatible.  Every symplectic manifold admits 
an almost complex structure $J$, and a K\"ahler
metric $g$, such that $(g,J,\omega)$ is compatible.

\begin{prop}\label{comptripprp}
For any symplectic manifold $(M,\omega)$, together with a
symplectic distribution $V$, there 
exists a compatible triple $(g,J,\omega)$,  such that
$\PV$ is symmetric with respect to $g$, and $\PV JX=J\PV X$
for all $X\in\Gamma(TM)$.
\end{prop}

\begin{proof}
We pick a smooth Riemannian metric $\tilde{g}$ on $M$,
relative to which $\PV$ is symmetric, for instance by choosing 
an arbitrary Riemannian metric $g'$ on $M$,
and defining
$\tilde{g}(X,Y):= g'(\PV X,\PV Y)+g'(\bar{\pi}_V X,\bar{\pi}_V Y)$,
where $\bar{\pi}_V=\1-\PV$.
We consider the non-degenerate, smooth bundle map $K$ defined by
$\omega(X,Y)=\tilde{g}(K X,Y)$, which is skew symmetric with respect to 
$\tilde{g}$, that is, its 
$\tilde{g}$-adjoint is $K^*=-K$. 
$K^* K=-K^2$ is  smooth, positive definite, non-degenerate and
$\tilde g$-symmetric, hence
there is a unique smooth, positive definite, $\tilde g$-symmetric bundle map $A$
defined by
$A^2=-K^2$,  which commutes with $K$.  Consequently, the
bundle map $J=KA^{-1}$ satisfies
$J^2=-\1$, and defines an almost complex structure. 
Since $A$ is positive definite and $\tilde g$-symmetric,  
$g(X,Y):=\tilde{g}(AX,Y)$ is a Riemannian metric with 
$\omega(X,Y)=g(JX,Y)$.
Moreover, this metric is hermitean, since
$g(JX,JY) = \tilde{g}(KX, A^{-1}K Y) 
        =  -\tilde{g}(X,K^2 A^{-1}Y)  
        = \tilde{g}(X,AY) 
        = g(X,Y)$.
In fact, since $\omega$ is closed, $g$ is  K\"ahler.

To show that  
$\PV$ is $g$-symmetric, we note that
$\PV$ commutes with $K$, since
$\tilde{g}(K\PV X,Y)=\omega(\PV X,Y)=\omega(X,\PV Y)=\tilde{g}(KX,\PV Y)=
      \tilde{g}(\PV KX,Y)$
for all $X,Y\in\Gamma(TM)$, using that $\PV$ is symmetric with
respect to $\tilde{g}$.
Hence, $\PV$ commutes with $A^2=-K^2$, and it straightforwardly follows from 
the $\tilde g$-orthogonality of $A$, $\PV$, and the positivity of $A$
that $\PV$ and $A$ commute. Hence, $\PV$ is $g$-symmetric, and it is also
clear that $\PV$ commutes with $J=KA^{-1}$. Thus, $J$
in particular restricts to a bundle map
$J:V\rightarrow V$. \end{proof}

\subsection{Further Properties} 

Some key properties of Hamiltonian systems concerning symmetries,
Poisson brackets, energy conservation, and, to some degree,
symplecticness, can be generalized to the constrained
system.

\subsubsection{ Symmetries}
Let us assume that the Hamiltonian system
$(M,\omega,H)$  admits a
symplectic $G$-action ($G$ some Lie group)
$\Psi:G\rightarrow {\rm Diff}(M)$, such that
$\Psi_h^*\omega=\omega$ and $H\circ\Psi_h=H$ for all $h\in G$.
Then,
we will say that the constrained system $(M,\omega,H,V)$ 
admits a $G$-symmetry
if $\Psi_{h*} V=V$ holds for all $h\in G$.

\subsubsection{Generalized Dirac bracket}
The smooth, $\R$-bilinear,
antisymmetric pairing on $C^\infty(M)$ associated to $(M,\omega,V)$ given by
\eqn\label{Vbra}\lbrace f,g\rbrace_V\;:=\;\omega(\PV(X_f),\PV(X_g))\;\eeqn
is a straightforward generalization of  the Dirac and Poisson brackets.
Along orbits of $\tPhi_t$, one has
$$\partial_t f(\tPhi_t(x))\;=\;\lbrace H,f\rbrace_V(\tPhi_t(x))\;$$
for all $x\in M$,
in analogy to (~\ref{Poisseqsmo}).
However, the bracket (~\ref{Vbra}) does not satisfy the Jacobi identity
if $V$ is non-integrable, but it
satisfies a Jacobi identity on every
(symplectic) integral manifold if $V$ is integrable.

\subsubsection{ Energy conservation} 
This key conservation law also holds for the constrained system.

\begin{prop}
The energy $H$ is an integral of motion of the dynamical system
(~\ref{eqsofmo}).
\end{prop}

\begin{proof}
This follows from the antisymmetry of the generalized Dirac bracket,
which implies that $\partial_t H=\lbrace H,H\rbrace_V=0$. \end{proof}

\subsubsection{Symplecticness}
The flow $\tPhi_t$ is not symplectic, but the following holds. Let us consider
\eqnn\partial_t\tPhi_t^{*}\omega&=&\tPhi_t^* \cL_{X_H^V}\omega 
       =  \tPhi_t^*\big(d i_{X_H^V}\omega + i_{X_H^V}d\omega\big) \\
       &= & -\;\tPhi_t^*\;d\left(\;(\PV)^i_k \;\partial_i H \; dx^k\;\right) 
       =  - \tPhi_t^*\;\left(\;\partial_l(\PV)^i_k\;\partial_i H\;
             dx^l\wedge dx^k\;\right)\\
      &=& - \frac{1}{2} \tPhi_t^* 
           \Big(  \big( \partial_l(\PV)^i_k -
           \partial_k (\PV)^i_l \big) 
           \partial_i H dx^l \wedge dx^k \Big)\;.\eeqnn
Hence, the restriction of
$\partial_t\tPhi_t^*\omega$ to $X,Y\in\Gamma(V)$ is given by
\eqnn\partial_t\tPhi_t^*\omega(X,Y)\;=\;
      - \frac{1}{2} \tPhi_t^* 
       \Big( 
       \cF^i_{rs} \partial_i H\,X^r Y^s \Big)
       \;,\eeqnn
where $\cF^i_{rs}$ is defined in lemma {~\ref{Frob}}. Thus, 
the right hand side vanishes identically if and
only if $V$ is integrable.
In the latter case,  the restriction of $\tPhi_t^*\omega$ to $V\times V$
equals its value for $t=0$, given by
$\omega(\PV(\cdot),\PV(\cdot))$. On every
integral manifold $j:M'\rightarrow M$ of $V$,
$\tPhi_t$ is symplectic with respect
to the pullback symplectic structure $j^*\omega$.

\section{\nbf THE GEOMETRY AND TOPOLOGY OF THE  
CRITICAL MANIFOLD}\label{sectionII}

In this main section, we address geometrical and
topological properties of
the critical set $\crit$ of the constrained Hamiltonian system $(M,\omega,H,V)$.
The main result of the subsequent analysis is, for $M$ compact and
without boundary, the topological formula (~\ref{Conley-Zehnder0}) that interrelates 
the Poincar\'e polynomials of $M$ and  
$\cg$ in a manner closely akin to the Morse-Bott inequalities.
This result implies that the 
topology of $M$ necessitates the existence of certain 
connectivity components of $\cg$
of a prescribed index.  
The analysis is structured as follows.   

In section {~\ref{genpropsubsec}}, we prove that $\crit$ is,
in the sense of Sard's 
theorem, generically a smooth $2(n-k)$-dimensional submanifold 
$\cg\subset M$. For the special case in which $V$ is integrable, 
it is shown that
the intersection of any integral manifold of $V$ with $\cg$ is
a discrete set, in agreement with the usual understanding that critical
points in Hamiltonian systems - on every leaf of the
foliation in the integrable case -  are typically isolated.
  
In section {~\ref{topcritmfsubsec}}, we introduce the main tool
for the analysis of $\crit$, an auxiliary
gradient-like flow $\phi_t\in{\rm Diff}(M)$ 
generated by the vector field $\PV\grad H$, where
$g$ is the K\"ahler metric of the compatible quadruple 
introduced after proposition
{~\ref{comptripprp}}. From this point on, we assume that 
$H:M\rightarrow\R$ is a Morse function.
Let 
$j:\cg\hookrightarrow M$
denote the embedding.  
We show that $\cg$ is normal hyperbolic with 
respect to $\phi_t$, and that the critical points of  
$j^*H$ on $\cg$ are precisely those of $H$ on $M$.
The latter is quintessential for our discussion of the topology of
$\cg$ via comparison of the Morse-Witten complexes of $(\cg,j^*H)$
and of $(M,H)$ in section {~\ref{mowiproofsubsec}}.

In section {~\ref{cozeproofsubsec}}, we prove  
(~\ref{Conley-Zehnder0}) by an application of Conley-Zehnder theory, 
\cite{CoZe}, to the auxiliary gradient-like
system. The assumptions on $\crit$ are slightly less strict than
genericity. In particular, assuming 
that $\crit\setminus \cg$ is a disjoint union
of $C^1$ manifolds, 
we show that every connectivity component of $\crit\setminus \cg$
is contained in a
$H$-level surface, and that $\crit\setminus \cg$ can be deformed away by an 
infinitesimal perturbation of the vector field.

In section {~\ref{mowiproofsubsec}}, we assume that $\crit=\cg$, and
give a second proof based on the comparison of the
Morse-Witten complexes of $(M,H)$ and $(\cg,H|_{\cg})$.
Our construction only uses the theory 
for non-degenerate Morse functions, not for Morse-Bott functions.
The interest in this discussion is to elucidate the relationship between
critical points of the 'free' system $(M,H)$, and of the critical manifold
$\cg$ of the constrained system $(M,H,V)$.  
The special case of
mechanical systems (where $M$ is noncompact) will be analyzed in a
later section.

\subsection{\nbf Generic Properties of the Critical Set}
\label{genpropsubsec}

Let us to begin with recall some basic definitions.
Critical points of $H$ are given by zeros of $dH$,
and a corresponding value of $H$ is called a critical value. 
A critical level surface $\Sigma_{E}$ corresponds to a critical value $E$ of $H$,
whereas a regular level surface $\Sigma_E$ contains no
critical points of $H$ (the corresponding value of
$E$ is then called regular).
The critical set of the constrained Hamiltonian system
$(M,\omega,H,V)$ is given by
$$\crit\;=\;\left\lbrace x\in M\;|\;X_H^V(x)\;=\;0\right\rbrace\;
    \;\subset\;\;M\;.$$
The following theorem holds independently
of the fact whether $V$ is integrable or not.

\begin{thm}\label{Sardthm}
In the generic case, the critical set is a piecewise
smooth, $2(n-k)$-dimensional submanifold of $M$.
\end{thm}

\begin{proof}
Let $\lbrace Y_i\rbrace_{i=1}^{2k}$ denote a smooth, local family of
spanning vector fields for $V$ over an open neighborhood $U\subset M$.
Since $V$ is symplectic, the fact that $X_H^V$ is a section of $V$
implies that $\omega(Y_i,X_H^V)$ cannot be identically zero for all $i$
and everywhere in $U$.
Due to the $\omega$-skew orthogonality of $\PV$, and $\PV Y_i=Y_i$,
$$\omega(Y_i,X_H^V)\;=\;\omega(\PV (Y_i),X_H)\;=\;\omega(Y_i,X_H)
    \;=\;Y_i(H)\;.$$
Thus, with $\underline{F}:=(Y_1(H),\dots,Y_{2k}(H))\in
   C^\infty(U, \R^{2k})$,
it is clear that $\crit\cap U=\underline{F}^{-1}(\underline{0})$.
Since $\underline{F}$ is smooth, Sard's theorem implies
that regular values, having smooth, $2(n-k)$-dimensional
submanifolds of $U$ as preimages,
are dense in $\underline{F}(U)$ \cite{Mi2}. \end{proof}

For future technical convenience, we pick a local spanning
family  $\left\lbrace Y_i\in\Gamma(V)\right\rbrace_{i=1}^{2k}$
for $V$ that  satisfies
$$\omega(Y_i,Y_j)\;=\;\tJ_{ij}\;,$$
with
$\tJ:=\left(\begin{array}{cc}0&\1_k\\-\1_k&0\end{array}\right)$.
This choice is always possible.

Furthermore, introducing the
associated family of 1-forms $\lbrace\theta_i\rbrace$
by $\theta_i(\cdot):=\omega(Y_i,\cdot)$,
theorem {~\ref{Pconstr}}
implies that
$$\PV\;=\;\tJ^{ij}Y_i\otimes \theta_j\;,$$
where $\tJ^{ij}$ are the components of $\tJ^{-1}=-\tJ$.
Expanding $X_H^V$ with respect to the basis $\lbrace Y_i\rbrace$ gives
\eqn\label{XHVexp} X_H^V\;=\;\PV(X_H^V)\;=\;-\;Y_i(H)\;\tJ^{ij}\; Y_j\;,\eeqn
where one uses the relationship $\theta_j(X_H^V)=Y_j(H)$
obtained in the proof of theorem {~\ref{Sardthm}}. Then, the following
proposition holds, which is in the subsequent discussion interpreted as
the property of normal hyperbolicity of $\crit$ with respect
to a certain gradient-like flow if the genericity
assumption is satisfied.

\begin{prop}\label{invert}
Under the genericity assumption of theorem {~\ref{Sardthm}},
the $2k\times 2k$-matrix given by
$[Y_k(Y_i(H))(\xa)]$ is invertible for all  $\xa\in\crit$,
and every local spanning family $\lbrace Y_i\in\Gamma(V)\rbrace$ of $V$.
\end{prop}

\begin{proof}
Let us pick a local
basis $\lbrace Y_i\rbrace_1^{2k}$ for $V$, and
$\lbrace Z_j\rbrace_1^{2(n-k)}$ for $V^\perp$, which together span $TM$.
Let $\xa\in\cg$, and assume the generic situation of theorem
{~\ref{Sardthm}}. Because $\cg$ is defined as the set of
zeros of the vector field (~\ref{XHVexp}), the kernel of the linear map
$$\left. dF_i(\cdot)\tJ^{ik}Y_k\right|_{\xa}\;:\;
       T_{\xa}M\;\rightarrow\;V_{\xa}\;,$$
where $F_i:= Y_i(H)$, is precisely $T_{\xa}\cg$,
and has a dimension $2(n-k)$.

In the basis $\lbrace Y_1|_{\xa},\dots,Y_{2k}|_{\xa},Z_1|_{\xa},\dots,
Z_{2(n-k)}|_{\xa}\rbrace$, its  matrix is given by
$$A\;=\;\left[ A_V\;A_{V^\perp}\right]\;,$$
where $A_V:=[Y_i(F_j)\tJ^{jk}Y_k|_{\xa}]$, and
$A_{V^\perp}:=[Z_i(F_j)\tJ^{jk}Y_k|_{\xa}]$.
Bringing $A$ into upper triangular form, $A_V$ is likewise 
transformed into upper
triangular form. Because the rank of $A$ is $2k$, and $A_V$ is a
$2k\times 2k$-matrix, its upper triangular form has $2k$
nonzero diagonal elements. Consequently, $A_V$ is invertible,
and due to the invertibility of $\tJ$, one arrives at the
assertion. \end{proof}

\begin{cor}\label{SpecQaminPalm}
Let $\{Y_i\}_{i=1}^{2k}$ denote a local spanning family for $V$,
and let $\{X_i\}_{i=1}^{2k}$ be a local spanning family (of $C^\infty$
sections) of 
$N\cg$, interpreted as a vector bundle over $\cg$ that is embedded
in $\cup_{x\in\cg}T_xM$. Then, 
the matrix $[g(Y_i,X_j)(x)]_{i,j=1}^{2k}$ is invertible for every $x\in\cg$.
\end{cor}

\begin{cor}\label{intcasecor}
Let $\cg$ satisfy the genericity assumption of theorem {~\ref{Sardthm}}.
If $V$ is integrable, the intersection of any integral manifold
of $V$ with $\cg$ is a discrete set.
\end{cor}

\begin{proof}
The previous proposition implies that generically, integral manifolds of
$V$ intersect $\cg$ transversely.
Their dimensions are mutually
complementary, hence the intersection set is zero-dimensional. \end{proof}

\subsection{Normal Hyperbolicity and an Auxiliary Gradient-Like System}
\label{topcritmfsubsec}
In this section, we introduce the main tool necessary for the analysis
of the topology of $\crit$, given by an auxiliary gradient-like flow
on $M$ whose same critical set is also $\crit$.
Furthermore, we show that generically, $\crit$ is normal hyperbolic
with respect to the latter.

\subsubsection{A Generalized Hessian}

To begin with, we define a generalized Hessian for $\crit$.
The usual coordinate-free definition of the Hessian of $H$ is
$\nabla dH$, evaluated at
the critical points of $H$, where $\nabla$ is the Levi-Civita connection
of the K\"ahler metric $g$. 
Let  $\PV^\dagger$ denote the dual projection tensor associated to $\PV$, 
which acts on sections of the
cotangent bundle $T^*M$, so that for any 1-form $\theta$,
$\langle \PV^\dagger\theta, X\rangle =\langle\theta,\PV X\rangle$.
The generalization of the Hessian in our context is the tensor
$\nabla (\PV^\dagger dH)$, which acts as a bilinear form on $\Gamma(TM)\times\Gamma(TM)$
by way of
\eqnn\nabla(\PV^\dagger dH)(X,Y)&:=&\langle\nabla_X (\PV^\dagger dH),
         Y\rangle \\
         &=&(((\PV)^j_r H_{,j})_{,s} -\Gamma^s_{ri}(\PV)_s^j H_{,j} ) 
         X^r Y^s ,\eeqnn
where $\Gamma^s_{ri}$ are the Christoffel symbols.
Evaluating this quantity on $\crit$,
the second term in the bracket on the lower line is zero.
The non-vanishing term is determined by the matrix
\eqn\label{hessgen}[K_{rs}] := [((\PV)^j_r H_{,j})_{,s}] .\eeqn
One straightforwardly verifies that
$(\PV)^j_i  K_{jk} = K_{ik}$ is satisfied
everywhere on $\crit$, hence rank$\{K\}\leq$ rank$\{\PV\}=2k=$rank$\{V\}$.
Clearly, the corank of
$K|_a$ equals the dimension of the connectivity component of $\crit$
containing $a$.

\subsubsection{Definition of the Gradient-Like System}
\label{auxgradlkflowsssec}

A flow is gradient-like if there exists
a function $f:M\rightarrow\R$ that decreases strictly along all of its
non-constant orbits.
The flow $\tPhi_t$ of the constrained Hamiltonian system is not gradient-like,
and hence
turns out to be of limited use for the study of the global
topology of $\crit$, because invariant sets of $\tPhi_t$ 
do in general not only contain fixed points, but also
periodic orbits.

Instead, we introduce the auxiliary dynamical system  
\eqn\label{gradlikeODE}\partial_t\gamma(t)
    =-(\PV \nabla_g H  )(\gamma(t))\;, \eeqn
where $\gamma:I\subset\R\rightarrow M$,
which turns out to be an extremely powerful tool for our purpose.
Let us denote its flow by
$\phi_t\in{\rm Diff}(M)$.
The orbits of (~\ref{gradlikeODE}) are clearly $V$-horizontal,
and both $\Phi_t^c$ and $\phi_t$
exhibit the same critical set $\crit$.

\begin{prop}
The flow $\phi_t$ is gradient-like.
\end{prop}

\begin{proof}
Since
\eqnn\partial_t H(\gamma(t))&=&\left\langle dH(\gamma(t))\, , \,
     \partial_t \gamma(t) \right\rangle =
		- g\left(\grad H\, , \, \PV\grad H\right)(\gamma(t))\\
     &=&- g\left(\PV\grad H\, , \,\PV\grad H\right)(\gamma(t)) 
		\leq  0 ,\eeqnn
it follows that $H$ decreases strictly along the
non-constant orbits of $\phi_t$.
We have here used the fact that $(g,J,\omega,\PV)$ is a compatible quadruple.
\end{proof}

$(g,J,\omega,\PV)$ has been constructed for this precise reason.
It is immediately clear that
$\phi_t$ generates no periodic trajectories,
hence $\crit$ comprises all invariant sets of $\phi_t$.


\subsubsection{\nit Morse Functions and Non-Degenerate Critical Manifolds}
 
Let us next recall some standard definitions from
Morse- and Morse-Bott theory that will be needed in the subsequent discussion.
The dimensions of the  zero and negative eigenspaces of the Hessian  of
$f$ at a critical point $a$ are
called the nullity and the index of the critical point $a$.
If all critical points of $f:M\rightarrow\R$ have a zero nullity,
$f$ is called a Morse function, and the index is
then called the Morse index of $a$.
If the critical points of $f$ are not isolated, but elements of
critical manifolds that are non-degenerate in the sense of Bott,
$H$ is called a Morse-Bott function \cite{Bo}.
Throughout this section, we will
assume that $H$ is a Morse function.

Furthermore, we recall some standard definitions related to normal
hyperbolicity, applied to the case of $\crit$.
 A connectivity component $\crit_i$ is locally
normal hyperbolic at the point $a\in\crit$
with respect to $\phi_t$ if it is a manifold at $a$, and
if the restriction of $\K$ to the normal space $N_a\crit$ is non-degenerate.
A connectivity component $\crit_i$ is called non-degenerate
if it is a manifold that is everywhere normal hyperbolic with respect
to $\phi_t$.
The index of a non-degenerate
connectivity component $\crit_i$ is
the number of eigenvalues of the constrained Hessian
$\K$ on $\crit_i$ that are contained in the negative half plane.

\begin{prop}\label{normhypgenprop}
If $\crit$ is generic in the sense of Sard's theorem, 
it is normal hyperbolic with
respect to the gradient-like flow $\phi_t$.
\end{prop}

\begin{proof} This follows straightforwardly from proposition {~\ref{invert}}. \end{proof}

Let $\crit_i$, $i=1,..,l$ denote the connectivity
components of $\crit=\cup \crit_i$, and let $j_i:\crit_i
\hookrightarrow M$ denote
the embedding of the $i$-th components.

\begin{prop}\label{critMorseH1}
Assume that $\crit$ satisfies the genericity assumption in the sense of
Sard's theorem, and that $H:M\rightarrow\R$ is a Morse function. Then,
$H_i:=H\circ j_i:\crit_i\rightarrow\R$ is a Morse function, and 
$x\in \crit_i$ is a critical point
of $H_i$ if and only if it is a critical point of $H$.
\end{prop}

\begin{proof}
It is trivially clear that every critical point of $H$
is a critical point of $H_i$.
To prove the opposite direction, suppose
that $a$ is an extremum of $H|{\crit_i}$.
Then, $\grad H|_a\in N_a\crit$, but also, by definition of $\crit$,
$P_a\grad H|_a = 0$. By corollary {~\ref{SpecQaminPalm}},
this can only be true if $\grad H_a=0$.  
The Hessian of the restriction of $H$ at any critical point of $H_i$
is nondegenerate, thus $H_i$ is 
a Morse function on $\crit_i$. \end{proof}

\begin{cor}\label{critMorseH2}
The critical points of $H|_{\cg}:\cg\rightarrow\R$ are precisely the 
critical points of $H:M\rightarrow\R$. If $\crit_i$ is a
non-generic
connectivity component  
that is a normal hyperbolic submanifold of $M$,
$\crit_i\subset\Sigma_{H(\crit_i)}$.
\end{cor}

\begin{proof} The first assertion follows trivially from the previous
proposition.
Assuming that $\crit_i$ is a non-generic connectivity
component of $\crit$ that is a manifold and normal hyperbolic, 
the previous
proposition implies
that there are no extrema of $H|_{\crit_i}$. Thus,
$\crit_i$ is a submanifold of the level surface
$\Sigma_{H(\crit_i)}$. \end{proof}

\subsection{\nit Approach via Conley-Zehnder Theory}
\label{cozeproofsubsec}

The goal in this and the next section is to derive the relationship
(~\ref{Conley-Zehnder0}) between the Poincar\'e polynomials of $\crit$,
and $M$. We first approach this problem by use of 
Conley-Zehnder theory, \cite{CoZe}, and under slightly less 
restrictive assumptions than genericity. 

Let us for convenience first recall some of the 
key elements from Conley-Zehnder theory, \cite{CoZe}. 
Let $\crit_i$ be any compact component of $\crit$.
An index pair associated to $\crit_i$ is a pair of compact sets
$(N_i,\tilde{N}_i)$ that possesses the following properties.
The interior of $N_i$  contains $\crit_i$, and moreover,
$\crit_i$ is the maximal invariant set under $\phi_t$ in the interior of $N_i$.
$\tilde{N}_i$ is a compact subset of $N_i$ that has empty intersection with
$\crit_i$, and the trajectories of all points in $N_i$ that leave
$N_i$ at some time under the gradient-like flow $\phi_t$ intersect
$\tilde{N}_i$. $\tilde{N}_i$ is called the exit set of $N_i$.

The homotopy type of the pointed space
$N_i/ \tilde{N}_i$ only depends on $\crit_i$, by a result
proven in \cite{CoZe}, so that the relative cohomology
$H^*(N_i,\tilde{N}_i)$ (with coefficients appropriately chosen) 
is independent of the particular choice of index pairs
(the space $N_i/ \tilde{N}_i$ is obtained from collapsing the subspace
$\tilde{N}_i$ of $N_i$ to a point). The equivalence class
$[N_i/ \tilde{N}_i]$ of pointed topological spaces under homotopy
is called the {\em Conley index} of $\crit_i$.

Let $I$ denote a compact invariant invariant set under $\phi_t$.
A  Morse decomposition of
$I$ is a finite, disjoint family of compact, invariant subsets
$\lbrace M_1,\dots,M_n\rbrace$ that satisfies the following requirement on
the ordering. For every $x\in I\setminus \cup_i
M_i$, there exists a pair of indices $i<j$, such that
$\lim_{t\rightarrow-\infty}\phi_t(x)\subset M_i$, and 
$\lim_{t\rightarrow \infty}\phi_t(x)\subset M_j$.
Such an ordering, if it exists, is called admissible, and the $M_i$ are
called Morse sets of $I$.

For every compact invariant
set $I$ admitting an admissibly ordered Morse decomposition, there exists
an increasing sequence of compact sets $N_i$ with $N_0\subset
N_1\subset\dots\subset N_m$, such that $(N_i,N_{i-1})$ is an index pair for
$M_i$, and $(N_m,N_0)$ is an index pair for $I$, \cite{CoZe}.

Consider compact manifolds $A\supset B \supset C$. It
is a standard fact that the exact sequence of relative cohomologies
$$\dots\stackrel{\delta^{k-1}}{\rightarrow} H^k(A,B)\rightarrow
       H^k(A,C)\rightarrow H^k(B,C)      \stackrel{\delta^k}{\rightarrow}
       H^{k+1}(A,B)\rightarrow\dots$$
implies that, with $r_{i,p}$ denoting the rank of $H^p(N_i,N_{i-1})$, 
$$\sum_{i,p}\lambda^p r_{i,p}=\sum_p b_p \lambda^p + (1+\lambda)\Q(\lambda) $$
for the indicated Poincar\'e polynomials (cf. for instance \cite{Jo}).
$b_j$ is the $j$-th Betti number of the index pair $(N_m,N_0)$ of $I$, and
$\Q(\lambda)$ is a polynomial in $\lambda$ with non-negative integer coefficients.
Due to the positivity of the coefficients of
$\Q(\lambda)$, it is clear that $\sum_i r_{i,p}\geq b_p$.

If $M$ is  compact and  closed, and if $\crit$
is non-degenerate, the following holds. The invariant set $I$ can be
chosen to be equal to $M$. We let 
$N_m=M$ and $N_0=\emptyset$ denote the top and bottom elements of the
sequence defined above, and order the connected elements of
$\crit$
according to the descending values of the maximum of $H$ attained on each
$\crit_i$. Then, $\crit$ furnishes a Morse decomposition for
$M$. The homology groups of $M$ are isomorphic to the relative homology
groups of the index pair $(N_m,N_0)$. Hence the numbers
$b_p$ are the Betti numbers of $M$.

\begin{prop}
Let $\crit_i\subset\cg$ be a generic connectivity component, 
compact and without boundary, and let
$(N_i,\tilde{N}_i)$ denote any associated index pair. Then,
\eqn H^{q+\mu_i}(N_i,\tilde{N}_i)\;\cong\; H^q(\crit_i)\;,\eeqn
where $\mu_i$ is the index of $\crit_i$, and $q=0,\dots,dim(\crit_i)$.
\end{prop}

\begin{proof}
We consider, for  $\epsilon_0>0$ small,
a compact tubular
$\epsilon_0$-neighborhood $U$ of $\crit_i$ (of dimension $2n$), and let
$$W_U^{cu}(\crit_i)\;:=\;(W^-(\crit_i)\cup \crit_i)\cap U$$
denote the intersection of the center unstable manifold of $\crit_i$
with $U$. $W^-(\crit_i)$ denotes the unstable manifold of $\crit_i$.
Pick some small, positive $\epsilon <\epsilon_0$, and let $U_\epsilon$
be the compact tubular $\epsilon$-neighborhood
of $W_U^{cu}(\crit_i)$ in $U$.

Letting $\epsilon$ continuously go to zero, we obtain
a homotopy equivalence of tubular neighborhoods,
for which $W_U^{cu}(\crit_i)$ is a deformation retract. Let
$$U_\epsilon^{out}\;:=\;\partial U_\epsilon \cap \phi_\R(U_\epsilon)$$
denote the intersection of $\partial U_\epsilon$ with all orbits of the
gradient-like flow that contain points in $U_\epsilon$.
Then, $(U_\epsilon,U_\epsilon^{out})$ is an index pair for
$\crit_i$, and by letting $\epsilon$ continuously go to zero,
$U_\epsilon^{out}$ is homotopically retracted to
$\partial W_U^{cu}(\crit_i)$.

Thus, by homotopy invariance,
\eqnn H^*(U_\epsilon,U^{out}_\epsilon)\cong
      H^*(W_U^{cu}(\crit_i),\partial W_U^{cu}(\crit_i)).\eeqnn
Since $\crit_i$ is normal hyperbolic with respect to the
gradient-like flow, $W_U^{cu}(\crit_i)$ has a constant dimension
$n_i+\mu(\crit_i)$ everywhere, where $n_i={\rm dim}\crit_i$.
Therefore, by Lefschetz duality, \cite{DuFoNo},  
$$H^{n_i+\mu_i-p}(W_U^{cu}(\crit_i),\partial W_U^{cu}(\crit_i))\cong
        H_p(W_U^{cu}(\crit_i)\setminus\partial W_U^{cu}(\crit_i)) ,$$
where $\mu_i=\mu(\crit_i)$, the index of $\crit_i$.
Since $\crit_i$ is a deformation retract of the interior of
$W_U^{cu}(\crit_i)$, the
respective cohomology groups are isomorphic.

Due to $dim(\crit_i)=n_i$, we have by Poincar\'e duality  
$$H_p(W_U^{cu}(\crit_i)\setminus\partial W_U^{cu}(\crit_i))
      \cong H_p(\crit_i)\cong H^{n_i-p}(\crit_i),$$
so that with $q:=n_i-p$,
\eqn\label{holy}H^{q+\mu_i}(U_\epsilon,U^{out}_\epsilon)
    \cong H^q(\crit_i) ,\eeqn
which proves the claim.
\end{proof}

From (~\ref{holy}), we deduce that
$r_{i,p} = {\rm dim} H^{i,p-\mu_i}(\crit_i) $
(recalling that $\mu_i$ is the index of $\crit_i$), hence
$r_{i,p}=b_{p-\mu_i}(\crit_i)$. 
Assuming that the number of connected components of
$\crit$ is finite, one thus obtains
\eqn\label{Conley-Zehnder0}\sum_{i,p} \lambda^{p+\mu_i}b_{p}(\crit_i)= 
     \sum_p  \lambda^p b_p(M) +
     (1+\lambda) \Q(\lambda) ,\eeqn
which in particular implies  
$\sum_{i} b_{p-\mu_i}(\crit_i) \geq b_{p}(M)$.
Setting $\lambda=-1$, 
$$\sum_{p} (-1)^{p+\mu_i} b_{p}(\crit_i) = \sum_{i} (-1)^{\mu_i} \chi(\crit_i)
        = \chi(M) ,$$
where $\chi$ denotes the Euler characteristic.
\\

\noindent{\bf Remark.} In the case of mechanical systems, the phase space
of the relevant constrained Hamiltonian system is
non-compact,
and the critical manifold is generally unbounded.
Therefore, the arguments used here do not apply.
However, since in that case,
$M$ and $\crit$ are vector bundles, we are nevertheless able
to prove results that are fully analogous to (~\ref{Conley-Zehnder0}).
\\

We can prove a slightly more general result by relaxing 
the assumption of genericity.

\begin{prop}\label{Conley-Zehnderineqprp1nongeneric}
Assume that $\crit\setminus\cg$ is a disjoint union of $C^1$-manifolds. Then,
\eqn\label{Conley-Zehnder1}\sum_{\stackrel{i,p}{\crit_i\in\cg}}
       b_{i,p}\;\lambda^{p+\mu_i}\;=\;\sum_p \;b_p\;
      \lambda^p \;+\; (1+\lambda)\;\tilde{\Q}(\lambda) \;. \eeqn
$b_{i,p}$ are the $p$-th
Betti numbers of the connectivity components $\crit_i$ of
$\cg$, $b_p$ are the Betti numbers of $M$,
and $\tilde{\Q}$ is a polynomial with non-negative integer coefficients.
\end{prop}

\begin{proof}
We show that $X_H^V$ can be infinitesimally perturbed such that
$\crit\setminus\cg$ is removed.
Consider, for 
$\epsilon>0$, the compact tubular neighborhoods  
\eqn\label{neighb}U_\epsilon(\crit_i)=\lbrace x\in M\mid
      dist_g(x,\crit_i)\leq\epsilon\rbrace\eeqn
of connectivity components $\crit_i\subset \crit\setminus\cg$,
where $dist_g$ denotes the
Riemannian distance function
induced by $g$.
We introduce a vector field
$X_\epsilon$, given by $\PV\grad H$ in
$M\setminus U_\epsilon(\crit_i)$, and
in the interior of every $U_\epsilon(\crit_i)$ with
$\crit_i\subset\crit\setminus\cg$, by
\eqn\label{kiwi}X_\epsilon|_x=\PV\grad H|_x +\epsilon h(x)\grad
   H|_x\;.\eeqn
Here, 
$h\in C^1(U_\epsilon(\crit_i),[0,1])$, obeying
$h|_{\crit_i}=1$ and $h|_{\partial U_\epsilon(\crit_i)}=0$ 
is strictly monotonic along the
the flow lines generated by $\PV\grad H$.  
$h$ exists because $\crit\setminus\cg$ is a disjoint union
of $C^1$-manifolds.

For all $\crit_i\subset\crit\setminus\cg$,
$\grad H$ is strictly non-zero in $U_\epsilon(\crit_i)$, as shown above.
We have
$$g(X_\epsilon,\grad H)=(\|\PV\grad H\|_g^2)(x)+\epsilon h(x) (\|\grad
      H\|_g^2)(x) ,$$
where we have used the $g$-symmetry of $\PV$, and
$\|X\|_g^2\equiv g(X,X)$.
The first term on the right hand side is non-zero on
the boundary of $U_\epsilon(\crit_i)$, while the second term vanishes.
Moreover, the second term is non-zero everywhere in the interior of
$U_\epsilon(\crit_i)$. Therefore, $X_\epsilon$ vanishes nowhere in
$U_\epsilon(\crit_i)$.
Hence, $X_\epsilon$ is a deformation of
$\PV\grad H$, with critical set $\cg$.
Notably, $\cg$ cannot be removed in this manner,
since it contains critical points of $H$.

Since $g(X_\epsilon,\grad H)$ is strictly positive in every
$U_\epsilon(\crit_i)$,  $X_\epsilon$ generates a gradient-like
flow.
Since $\|X_\epsilon-\PV\grad H\|_g\leq O(\epsilon)$ everywhere on $M$,
one can pick $X_\epsilon$
arbitrarily close to $\PV\grad H$ in the $L^\infty$-norm on
$\Gamma(TM)$ induced by $\|\cdot\|_g$.
Carrying out
the Conley-Zehnder construction with respect to the flow generated by
$X_\epsilon$ yields (~\ref{Conley-Zehnder1}).
This result does not require the assumption of normal hyperbolicity
on $\crit$. \end{proof}

\subsection{\nbf Approach via the Morse-Witten Complex}
\label{mowiproofsubsec}

We will next provide a different derivation of (~\ref{Conley-Zehnder1}), 
based on the construction of the Morse-Witten differential complex.
The motivation is to clarify the orbit structure of the gradient-like 
system, and to devise an explicit construction that relates the 
Morse-Witten complexes of the free and constrained system to one
another. This in particular only involves the 
corresponding theory for {\em non}-degenerate  Morse functions.

Let us to begin with briefly recall the basic framework of this construction.
Let $M$ be a compact, closed, orientable and smooth $n$-manifold,
and let $f:M\rightarrow\R$ be a Morse function.
Let $\Cb^p$ denote the free $\Z$-module generated
by the critical points of $f$ with a Morse index $p$.
The set $\Cb=\oplus_p \Cb^p$ is the free $\Z$-module generated by the
critical points of $f$, and graded by their Morse indices.
There exists a natural coboundary operator
$\delta:\Cb^p\rightarrow\Cb^{p+1}$, with $\delta\circ\delta=0$, 
whose construction we recall next, cf. \cite{AuBr,Bo1,Fl,Wi}.

Introducing an auxiliary Riemannian structure on $M$, we
let $W_a^-$ and $W_a^+$ denote the unstable and stable manifold of the 
critical point $a$ of $f$ under the gradient flow, respectively, and assign an
arbitrary orientation to every $W_a^-$.
The orientation of $M$, together with the orientation of $W_a^-$
at every critical point $a$ induces an orientation of $W_a^+$.
Morse functions, for which all $W_a^-$ and $W_{a'}^+$ 
intersect transversely, are dense in $C^\infty(M)$.
The dimension of $W_a^-$ equals the Morse index $\mu(a)$ of $a$,
and the dimension of the intersection
$M(a,a'):= W_a^-\cap W_{a'}^+$
is given by  ${\rm max}\lbrace\mu(a)-\mu(a'),0\rbrace$.
For pairs of critical points $a$ and $a'$ with relative Morse index 
$1$, say $\mu(a)=\mu(a')+1$, $M(a,a')$ is a
finite collection of gradient lines that connect $a$ with $a'$.

The intersection of $M(a,a')$ with any regular level
surface $\Sigma_c$ of $f$ with $f(a)<f(\Sigma_c)=c<f(a')$ is transverse,
and consists of a  finite collection of isolated points.
Then, one picks the orientation of $\Sigma_c$, which, combined with the section
$\grad f$ of its normal bundle, shall agree with the orientation of $M$.
The submanifolds $W_{a,c}^-:= W_a^-\cap\Sigma_c$ and
$W_{a',c}^+:= W_{a'}^+\cap\Sigma_c$ of $\Sigma_c$ are smooth, compact and 
closed, with complementary dimensions in $\Sigma_c$, and orientations picked above.
Hence, their intersection number, which is often in this context written as
$\langle a,\delta a'\rangle := \sharp(W_{a,c}^-,W_{a',c}^+)$,
is well-defined,  \cite{Hi}.
The coboundary operator of the Morse-Witten complex is defined as the $\Z$-linear map
$\delta:\Cb^p\rightarrow\Cb^{p+1}$,  defined by
$$\delta a' = \sum_{\mu(a)=p+1} \langle a,\delta a'\rangle a .$$

\begin{thm}\label{CohWitdRhthm}
The cohomology of the differential complex $(\Cb,\delta)$
is isomorphic to the de Rham
cohomology of $M$,
${\rm ker}\delta / {\rm im}\delta \cong H^*(M,\Z)$.
\end{thm}

The proof   can for instance be found in
\cite{AuBr,Fl,Sw,Wi}. If $\langle a,\delta a'\rangle\neq0$ for a pair $a$ and
$a'$ of critical points with a relative Morse index $1$,
we will say that they are effectively connected (by gradient lines).

As is well-known, the existence of the Morse-Witten complex implies the strong
Morse inequalities in the following manner.
Let    $\Zb^p:={\rm ker}\delta\cap\Cb^p$ denote the $p$-th cocyle group,
$\B^p\subset\Cb^p$ the $p$-th coboundary group, and  
$\H^p:=\Zb^p\setminus\B^p$ the $p$-th cohomology group under $\delta$. Thus,
${\rm dim}\H^{p}=b_p(M)$ by theorem {~\ref{CohWitdRhthm}}. From
$${\rm dim}\Cb^p= b_p(M)+ {\rm dim}\B^p + {\rm dim}\B^{p+1} ,$$
where ${\rm dim}\Cb^p=N_p$ (the number of critical points of
$f$ with Morse index $p$), follows  
\eqn\label{oregon}
		\sum_{p=0}^n \lambda^p N_p  
     	=\sum_{p=0}^n \lambda^p b_p(M)+ (1+\lambda)\sum_{p=1}^n \lambda^{p-1}{\rm dim}\B^p  
\eeqn
(both $\B^0$ and $\B^{2n+1}$ are empty).
The coefficients of the 
polynomial  ${\mathcal Q}(\lambda)=\sum \lambda^{p-1}{\rm dim}\B^p$ are
evidently non-negative and integer.
Clearly, ${\rm dim}\B^p$ is the number of critical points of Morse index
$p$ that are effectively connected to critical points of Morse index
$p-1$ via gradient lines of $f$.

\subsubsection{\bf Comparing the Complexes for the Free and 
Constrained System}

The goal of our discussion here is to devise an explicit construction
that relates the Morse-Witten complex of the free system
$(M,H)$ to the one on the critical manifold $(\cg,H|_{\cg})$,
by a deformation of the gradient-like flow $\phi_t$. 
This will yield (~\ref{Conley-Zehnder1}).

Let  $\crit_i$ denote the $i$-th connectivity component of $\cg$, and
$\A_i:=\lbrace a_{i,1},\dots,a_{i,m}\rbrace$ the 
critical points of $H$ contained in $\crit_i$. Furthermore, let
$\mu(a_{i,r})$ be the associated Morse indices of $H:M\rightarrow\R$,
and $H_i:= H|_{\crit_i}$ denote the restriction of $H$ to $\crit_i$.
By proposition {~\ref{critMorseH2}} and corollary {~\ref{critMorseH2}}, 
$H_i:\crit_i\rightarrow \R$ is a Morse function,
whose critical points are precisely the elements of $\A_i$.
The index $\mu(\crit_i)$ of $\crit_i$ equals
the number of negative eigenvalues of the Hessian of $H$ at any $a_{i,r}\in\A_i$
whose eigenspace is normal to $\crit_i$.

The Morse index of $a_{i,r}$ with respect to $H_i$ is thus $\mu(a_{i,r})-\mu(\crit_i)$.
To define the Morse-Witten
complex associated to $\crit_i$, we introduce the free $\Z$-module
generated by the elements of $\A_i$,
graded by the Morse indices $p$ of
the critical points of $H_i$,
$$\Cb_i=\oplus_p \Cb^p_i.$$
To construct the coboundary operator 
$\delta_i:\Cb^p_i\rightarrow\Cb_i^{p+1}$,
one uses the gradient flow on $\crit_i$ generated by $H_i$, thus obtaining
\eqn{\rm ker}\delta_i / {\rm im}\delta_i \cong H^*(\crit_i,\Z) .\eeqn
Application of (~\ref{oregon}) shows that for every $\crit_i\in\cg$,
\eqn\label{ontario}\sum_{p} \lambda^p N_{i,p} &=&
         \sum_{p} \lambda^p b_p(\crit_i)+
        (1+\lambda)\sum_{p} \lambda^{p-1}{\rm dim}\B^p_i ,\eeqn
where $\B^p_i$ is the $p$-th coboundary group of the Morse-Witten complex of
$\crit_i$, and $N_{i,p}$ is the number of critical points of $H_i$ on
$\crit_i$ of Morse index $p$.

Since every critical point of $H$ lies on precisely one generic
component $\crit_i$, the number $N_q$ of critical points of $H$
with a Morse index $q$ is given by
$$N_p=\sum_{i}N_{i;p-\mu(\crit_i)} .$$
Thus, combining (~\ref{ontario}) with (~\ref{oregon}),
one obtains  
\eqnn\sum_{i,p;\crit_i\in\cg}\lambda^q
         b_{q-\mu(\crit_i)}(\crit_i)=\sum_q \lambda^q b_q(M)
         \hspace{3cm}\\
         +\;(1+\lambda)\sum_q \lambda^{q-1}
         \Big({\rm dim}\B^q-\sum_{\crit_i\in\cg}{\rm
         dim}\B_i^{q-\mu(\crit_i)}\Big)\;.\eeqnn
Hence, (~\ref{Conley-Zehnder1}) is equivalent to the statement that
the polynomial on the last line,
which is multiplied by $(1+\lambda)$, has
non-negative integer coefficients.

By a homotopy argument, we will now prove that for all $q$,
\eqn\label{wdder}{\rm dim}\B^q \geq \sum_{\crit_i\in\cg}{\rm
          dim}\B_i^{q-\mu(\crit_i)}\eeqn
holds, thus obtaining an alternative proof of (~\ref{Conley-Zehnder1}).
The main motivation here is to give an explicit construction that
geometrically elucidates 
this relation, noting that the left hand side is defined by the flow of 
the 'free' gradient-like system corresponding to 
$(M,\omega,H)$, while the right hand
side is defined by the constrained gradient-like system corresponding
to $(M,\omega,H,V)$.
We note that ${\rm dim}\B_i^{q-\mu(\crit_i)}$ denotes the number of critical
points of $H$ with a Morse index $q$ in $\crit_i$,
which are effectively connected to critical points of Morse
index $p+1$ in $\crit_i$ via
gradient lines of the Morse function $H_i$ on $\crit_i$.
Therefore, the sum on the right hand side of (~\ref{wdder}) equals the
number of those critical points of $H$ with a Morse index $q$, which are
effectively connected to critical points of Morse index $q+1$ via gradient
lines of the functions $H\circ j_i$ on all generic
$\crit_i$. Here, $j_i:\crit_i\rightarrow M$ denotes the corresponding
inclusion maps.

\subsubsection{ Proof of (~\ref{wdder}) }
Our strategy consists of constructing a  
homotopy of vector fields
$v_\sig $, with $\sig \in[0,1]$, 
whose zeros are hyperbolic and independent of
$\sig $, which generate gradient-like flows.
They interpolate between
$v_1:=\grad H$,  and  
$v_0$, which is a vector field that is tangent to  
$\cg$.
For every $\sig \in[0,1]$, we construct a coboundary
operator via the one-dimensional integral curves of $v_\sig $ that connect
its zeros.
These coboundary operators are independent of $\sig $, and act on the
free $\Z$-module $\Cb$ of the Morse-Witten complex associated to $(M,H)$.
(~\ref{wdder})
then follows from a simple dimensional argument.

\begin{lem}
There exists $v_0\in\Gamma(TM)$, which is gradient-like,
and tangent to $\cg$.
Furthermore, the zeros of $v_0$ are hyperbolic,
and identical to the critical points of $H$.
The dimension of any
unstable manifold of the flow generated by $-v_0$ equals the Morse
index of the critical point of $H$ from which it emanates.
\end{lem}

\begin{proof}
We recall the vector field
$X_\epsilon$ constructed in the proof of proposition
{~\ref{Conley-Zehnderineqprp1nongeneric}}, and
consider the compact tubular $\epsilon$-neighborhoods
$U_\epsilon(\cg)$, defined similarly as in (~\ref{neighb}).
Furthermore, let
$\bar{Q}=\bar{Q}^2$ (resp. $Q=\1-\bar Q$)
be $g$-orthogonal, smooth tensors on $TU_\epsilon(\cg)$
of fixed rank $2(n-k)$ (resp. $2k$), with ${\rm Ker}\{\bar Q(a)\}=N_a\cg$
(resp. ${\rm Ker}\{  Q(a)\}=T_a\cg$) for every $a\in\cg$.

We define $v_0$ as follows. In $M\setminus U_\epsilon(\cg)$,
it shall equal $X_\epsilon$,
and that for $x$ in $U_\epsilon(\cg)$,
it shall be given by
$$v_0(x):= (\PV\grad H)(x)+h(x)(\bar{Q}\grad H)(x) ,$$
where $h:U_\epsilon(\cg)\rightarrow[0,1]$ is a smooth function obeying
$h|_{\cg}=1$ and $h|_{\partial U_\epsilon(\cg)}=0$.
In particular, $h$ is assumed to
be strictly monotonic along all non-constant trajectories of the flow
generated by $\PV\grad H$, and $dh$ shall vanish on $\cg$.

It can be easily verified that
$v_0$ possesses all of the desired properties.
It generates a gradient-like flow, since outside of $U_\epsilon(\cg)$,
$g(\grad H,v_0)=g(\grad H,X_\epsilon)>0$,
as has been shown in the proof of proposition
{~\ref{Conley-Zehnderineqprp1nongeneric}}.
In the interior of $U_\epsilon(\cg)$, one finds
$g(\grad H,v_0)=\|\PV\grad H\|_g^2 +
      h\|\bar{Q}\grad H\|_g^2$,
due to the $g$-orthogonality both of $\PV$ and $\bar{Q}$.
The first term on the right hand side vanishes everywhere on
$\cg$, but at no other point in $U_\epsilon(\cg)$.
The second term equals $\|\bar{Q}\grad H\|_g^2$ on $\cg$.
Since evidently, $\bar{Q}\grad H|_{\cg}$ is the gradient field of the
Morse function
$H|_{\cg}:\cg\rightarrow\R$ relative to the Riemannian metric on $T\cg$
induced  by $g$, its zeros are precisely the critical points of $H$ on
$\cg$, and it possesses no other zeros.
Because $g(v_0,\grad H)>0$ except at the critical points of $H$, 
it is clear that $-v_0$
generates a gradient-like flow $\psi_{0,t}$,
so that $H$ is strictly decreasing along
all non-constant orbits.
Furthermore, it is clear from the given construction that
$v_0$ is tangent to $\cg$.

To prove the remaining statements of the lemma, we note that
the Jacobian matrix of $v_0$ at $a$ in a local chart is given by
\eqn Dv_0(a)= (D^2_a H)^\sharp +(\IPa P_a- \IQa Q_a)(D^2_a H)^\sharp  .
		\label{armgdd}\eeqn
There is no dependence on $h$ because $dh|_{\cg}=0$.
Furthermore, $(D^2_a H)^\sharp$ is defined
as the matrix $[g^{ij}H_{,jk}|_a]$ in the given chart,
and $P_a$ denotes the matrix of $\PV(a)$.
Normal hyperbolicity follows from the invertibility of $Dv_0(a)$,
which is verified in the proof of lemma {~\ref{SpecQaminPalm}} below.
\end{proof}

\begin{lem}
Let $v_\sig := \sig \grad H + (1-\sig )v_0$
with $\sig \in[0,1]$.
Then, the flow  $\psi_{\sig ,t}$ generated by $-v_\sig $ is
gradient-like for any $\sig \in[0,1]$.
The zeros of
$v_\sig $ are hyperbolic fixed points of $\psi_{\sig ,t}$,
and independent of $\sig $.
Thus, the dimensions of the corresponding unstable manifolds equal the Morse 
indices of the critical points of $H$ from which
they emanate, for all
$\sig $.
\end{lem}

\begin{proof}
We consider 
$g(\grad H,v_\sig )=\sig \|\grad H\|_g^2 +(1-\sig )g(\grad H,v_0)$.
The first term on the right hand side is obviously everywhere positive except
at the critical points of $H$, and the same has been proved previously for the
second term.
Thus, $H$ decreases strictly along all non-constant orbits of
$\psi_{\sig ,t}$, hence the latter is gradient-like.
The Jacobian of $v_\sig $ at a critical point of $H$ is given by
\eqnn Dv_\sig(a )=
		({\bf 1}_{2n}+(1-\sig )(\IPa P_a - \IQa Q_a))(D^2_a H)^\sharp . \eeqnn
$Dv_\sig(a )$ is invertible for all
$\sig \in [0,1]$, since $(D^2_a H)^\sharp$ is invertible,
and  ${\rm spec}\{\IPa P_a - \IQa Q_a\}\subset(-1,1)$.
To prove the latter, we first observe that ${\rm spec}\{\IPa P_a - \IQa Q_a\}\subset[-1,1]$ is trivial, because $P_a$ and $Q_a$ both have a spectrum $\{0,1\}$. 
$\{-1,1\}$ is not included, because otherwise,  
$\bar P_a Q_a$, respectively $P_a \bar Q_a$, would not have a full rank,
in contradiction to corollary {~\ref{SpecQaminPalm}}. \end{proof}

By smoothness of $v_\sig $,
it follows that $\psi_{\sig ,t}$ is $C^\infty$ in $\sig $.
Thus, $\sig$ smoothly
parametrizes a homotopy of stable and unstable manifolds of $\psi_{\sig ,t}$
emanating from the
critical points of
$H$.
Since the fixed points of $\psi_{\sig ,t}$ are independent of $\sig $,
and the dimensions of the corresponding unstable manifolds are equal to
the Morse indices of the critical points of $H$,
we consider, for every value of $\sig\in[0,1]$, the free $\Z$-module
$\Cb=\oplus_p \Cb^p$
that is generated by the critical points of $H$, and graded by their Morse
indices.
For every $\sig$, we define a coboundary operator on $\Cb$, using  
$\psi_{\sig ,t}$ as follows. Picking a pair of critical points of
$H$ with a relative Morse index $1$, we consider the unstable manifold
$W^-_{\sig ,a}$ of $a$, and the stable manifold
$W^+_{\sig ,a'}$ of
$a'$ associated to $\psi_{\sig ,t}$.
Since $\sig $ parametrizes a homotopy of such manifolds,
they naturally
inherit an orientation from the one picked for
$\sig =1$ in the construction of the coboundary operator
of the Morse-Witten complex for $(M,H)$.

Let $\Sigma_E$ denote a regular energy surface
for $H(a)<E<H(a')$.
$W_{\sig ,a}^\pm$ intersects 
$\Sigma_E$ transversely,
because $H$ is strictly decreasing along all
non-constant orbits generated by $-v_\sig $.
$W^-_{\sig ,a}\cap\Sigma_E$ and
$W^+_{\sig ,a'}\cap\Sigma_E$ define two homotopies of 
oriented submanifolds of $\Sigma_E$.
By homotopy invariance of their intersection number, 
the coboundary operators are independent of
$\sig $, and thus identical to the 
$\delta$-operator of the Morse-Witten complex given
for $\sig =1$.

The stable and unstable
manifolds of $\psi_{0,t}$ are
either confined to some $\crit_i$, or connect critical
points lying on different $\crit_i$'s.
Let us consider pairs of critical points of $H$ with a relative Morse
index $1$ that lie on the same component
$\crit_i\in\cg$, and the corresponding stable and unstable manifolds of
$\psi_{0,t}$ which are contained in $\crit_i$.
Since
$v_0|_{\crit_i}$ is the projection of $\grad H|_{\crit_i}$ to
$T\crit_i$, these stable and unstable manifolds are the same as those which
were used to define the Morse-Witten complex on
$(\crit_i,H_i)$.
Using only stable and unstable manifolds of $\psi_{0,t}$
contained in $\cg$, we construct an operator $\tilde{\delta}$
acting on $\Cb$ in the
same manner in which the coboundary operator was defined, thus obtaining
$\tilde{\delta}\equiv \oplus_i \delta_i $,
where $\delta_i$ is the coboundary operator of the Morse-Witten complex
associated to the pair $(\crit_i,H_i)$.
Let $P_i:\Cb\rightarrow\Cb_i$ stand for
the projection of the free $\Z$-module $\Cb$
generated by all critical points of $H$ to the one generated by
the critical points contained in $\crit_i$.
Eliminating all integral lines of $-v_0$ that connect critical points on
different connectivity components of $\cg$ in the above construction,
one sees that $\delta_i=P_i\delta P_i$, thus
$\tilde \delta=P_i\delta P_i$. Hence, clearly,
$${\rm dim} ({\rm im}\delta|_{\Cb^p})\geq{\rm dim}
     ({\rm im}\tilde{\delta}|_{\Cb^p})\;,$$
which precisely corresponds to (~\ref{wdder}). 
This completes the proof.

\section{\nbf QUALITATIVE ASPECTS RELATED TO  
CRITICAL STABILITY}
\label{sectionIII}

So far, we have established that in the generic case, the connectivity 
components of $\crit=\cg$ are embedded 
submanifolds of dimension $2(n-k)$ equal to the corank of $V$. 
Furthermore, we have seen that the topology of the symplectic manifold
$M$ enforces the existence of connectivity components of $\cg$ of  
certain prescribed indices with respect to the auxiliary gradient-like flow $\phi_t$.

In this section, we focus on the physical dynamics, characterized by the 
flow $\tPhi_t$ generated by $X_H^V$, of the 
constrained Hamiltonian system $(M,\omega,H,V)$ in a tubular $\epsilon$-neighborhood
of $\cg$, and particularly on the issue of stability.
Let $g$ again denote the
auxiliary K\"ahler metric introduced in section {~\ref{sectionII}}, with the 
induced Riemannian distance function given by $dist_R$ (in contrast to
the Carnot-Caratheodory distance function $d_{C-C}$
induced by $g$, which will also be considered).
We recall that a point $\xa\in\cg$ is stable if there
exists $\delta(\epsilon)>0$ for every $\epsilon>0$, so that for all $t$,
$dist_R(\tPhi_t(x),\xa)< \epsilon$
holds for all $x$ with $dist_R(x,\xa)<\delta(\epsilon)$.

To elucidate the key differences between the local dynamics in the vicinity of $\cg$
for the cases of integrable and non-integrable $V$, let us first
describe the situation where $V$ is integrable.
As proved in corollary {~\ref{intcasecor}}, $M$ is foliated into $2k$-dimensional
symplectic submanifolds which intersect $\cg$ transversely. Thus,
on every leaf ${\mathcal N}$, the equilibrium solutions are generically isolated points. 
Let the linear operator $\Omega_a$ correspond to the linearization of $X_H^V$ 
on $T_a M$ for some $a\in {\mathcal N}\cap\cg$, and restricted to the fibre 
$V_a=T_a {\mathcal N}\subset T_a M$. 
Its spectrum, if it is not purely imaginary, conclusively characterizes
the stability of $a$; we refer to this as the asymptotically (un)stable case. 
If the spectrum of $\Omega_a$ is purely imaginary, which we refer to as the critically stable case, is is well-known that if there exists a local Lyapunov function 
$L_{{\mathcal N}}: U(a)\cap{\mathcal N}\rightarrow\R$ for $a$, then $a$ is stable. 
 
If $V$ is non-holonomic, the situation is similar in the asymptotically
(un)stable case, but drastically different in the critically stable situation.
In the critically stable case, the presence of a local degenerate Lyapunov function
for a single equilibrium $a\in\cg$ is of limited use, since there is a whole submanifold
$\cg\cap\Sigma_E$ (the $H$-level set for the energy $E$ of the initial condition) of valid
equilibria for a given energy $E$. One may relax this condition to the existence of 
the following class of functions.

\begin{defn}\label{LDLFdef}
Let $\nabla_g^\perp$ denote the component of the gradient $\nabla_g$ normal
to $\cg$ with respect to $g$ at $\cg$. Let $U(a)$ be a $dist_R$-small open
neighborhood of $a\in\cg$.
A  local, degenerate, almost Lyapunov function for $a$ is a class $C^1$ function
$L:U(a)\rightarrow\R$, which satisfies $(\nabla_g^\perp L)(a')=0$ for all 
$a'\in\cg\cap U(a)$,
and $\|\nabla_g L\|_g>0$ for all $x\in U(a)\setminus \cg$. Furthermore,
$(\nabla_g dL)|_{a'}$ is positive definite quadratic form on $N_{a'}\cg$ 
for all $a'\in U(a)\cap\cg$,
and  $L(\tPhi_t(x_0))\leq L(x_0)$ for all $x_0\in U(x_0)$, and all $t$ such
that $\tPhi_t(x_0)\in U(a)$. 
\end{defn}

Notably, $L$ defined here is not a local degenerate Lyapunov function, because
$\crit_i\cap U(a)$ is not a critical level set (we remark that this would be 
equivalent to $L$ being a Morse-Bott function in $U(a)$), on which $L$ is extremal.

While the existence of $L$ guarantees that the orbit $\tPhi_t(x_0)$ remains within
a tubular $\epsilon$-neighborhood of $\cg\cap U(a)$ for all $t$ such that 
$\tPhi_t(x_0)\in U(a)$, it does not imply stability of $a\in\cg\cap U$. There is
an additional, necessary condition on the rational independence of the frequencies of the
oscillatory linear problem that must be imposed. Otherwise, an inner resonance, connected
to the appearance of small divisors, occurs, and $\tPhi_t(x_0)\in U(a)$ may evolve 
away from $a$, 
in a diffusive motion along the higher flag elements of 
$V$ that are approximately tangent to $\cg$, 
while along $V$, which is transverse to $\cg$, 
the motion is bounded and oscillatory. 

From the analysis in section {~\ref{sectionII}}, it 
is clear that for every connectivity component $\crit_i\subset\cg$
of index $\mu(\crit_i)=0$ (with respect to $\phi_t$), 
the Hamiltonian $H$ is a local degenerate,
almost Lyapunov function for all of its points. The minimum $a^*$ of 
$H|_{\crit_i}$ on $\crit_i$ is a local minimum of $H$, 
and hence stable (since $H$ serves as a Lyapunov function for $a^*$).
Hence, in particular, if $V$ is integrable, all points on $\crit_i$ are
stable if $\mu(\crit_i)=0$.
We also note that on the connectivity components $\crit_j$ with index
$\mu(\crit_j)>0$, $H$ is never a local degenerate, almost Lyapunov function.

The main focus in this section will be to discuss issues of this type. 
However, an essential part of sections {~\ref{avtheorsubsub}} and 
{~\ref{Loptsubsubsect}} will be in mathematically non-rigorous terms, since 
a rigorous treatment of the matters addressed there would fall into the domain of
KAM and Nekhoroshev theory, and is beyond the scope of 
the present work.  

A concrete aim in this discussion is to arrive at stability criteria for
equilibria of the constrained Hamiltonian system $(M,\omega,H,V)$.  
From an instructive, despite elementary, application of averaging theory, we 
conjecture a condition for the critically stable case 
that involves an incommensurability condition imposed on the frequencies 
of the linearized problem, as remarked above. 
In order to elucidate its geometric content, 
we study the dynamics in the vicinity of a critically stable equilibrium 
in a geometrically invariant form that is adapted to the
flag of $V$. Invoking a perturbation expansion based on this description,
we argue that the incommensurability condition, which might merely 
correspond to an artefact of the averaging method, cannot be omitted.
A rigorous proof of the conjectured stability criterion is 
left for future work.

\subsection{\nit Stability Criteria}

Let $\xa\in\cg$, and pick some small neighborhood
$U(a)\subset M$ together with an associated Darboux chart,
with its origin at $\xa$. The equations of motion are given by
$\partial_t x_t = P(x_t)\mcJ  H_{,x}(x_t)  =  X_H^V(x_t)$,
where the coordinates are given by
$x=(x^1,\dots,x^n,x_{n+1},\dots,x_{2n})$,
and $\mcJ $ is the symplectic standard matrix.
Furthermore, $H_{,x}$ abbreviates $\partial_{x}H$,
and $P$ is the $2n\times2n$-matrix representing the tensor
$\PV$. $\omega$-skew orthogonality of $\PV$ translates into
$P(x)\mcJ X(x)=\mcJ P^\dagger(x)X(x)$ for all vector
fields $X$.

\begin{prop}
There exists a chart in which the equations of motion have the form
\eqn
		\partial_t (y_t,z_t)=\big( \Al
     	y_t  + Y(z_t,y_t) \; , \; Z(z_t,y_t) \big)\;
		\in \R^{2k}\times\R^{2(n-k)}\; .\label{parttzZeq}\eeqn
In particular, $\Al$ corresponds to the restriction of $DX_H^V(0) \IP$
to $V_0$, and $|Y(y,z)|$, $|Z(y,z)|=O(|y|\,|z|)+O(|y|^2)$.
\end{prop}

\begin{proof}
In a sufficiently small vicinity $U\subset\R^{2n}$ of the origin 
(corresponding to $a$), one infers from corollary {~\ref{SpecQaminPalm}}
that $T_a\cg\oplus V_0= \R^{2n}$, for $a\in U\cap \cg$.
Accordingly, we choose local coordinates
$z\in U'(0)\subset\R^{2(n-k)}$  on $\cg$, and 
$\tilde y\in V_{0}$, noting that the decomposition
$x=a(z)+\tilde y$ for any $x\in U\subset\R^{2n}$ is unique, where
$a:\R^{2{n-k}}\hookrightarrow U$ is the (smooth) embedding. Let $y$ denote
the coordinates of $\tilde y$ with respect to some family of basis
vectors  for $V_0$.  Then, (~\ref{parttzZeq}) evidently follows
from Taylor expansion. \end{proof}

\subsubsection{Asymptotic (In)stability}

If spec$\{\Al\}\cup i\R=\emptyset$, there exists, by the center manifold
theorem, a coordinate transformation
$(y,z)\rightarrow(\bar{y},\bar{z})$, such that  (~\ref{parttzZeq}) becomes
\eqn\partial_t(\bar{y}_t  ,  z_t)=\big(\Al  \bar{y}_t
        + \bar{Y}(\bar{y}_t,\bar{z}_t)\;   , \; 0 \big)  \eeqn
\cite{ZeBlMa}, where $\bar{Y}(0,\bar{z})=0$ for all $\bar{z}$.
Thus, $\xa\in\cg$ is asymptotically unstable if
there are eigenvalues with a positive real part, and asymptotically stable if
all eigenvalues have a negative real part.
If $V$ is integrable, asymptotic stability is impossible,
because the dynamics is Hamiltonian on every integral manifold.
However, if $V$ is non-integrable, there is, to the author's knowledge, no
obstruction to the existence of asymptotically stable
equilibria, since the flow map is not symplectic.

\subsubsection{ An Elementary Application of Averaging Theory }
\label{avtheorsubsub}

In the case of critical stability, one has 
spec$\{\Al\}=\lbrace i\omega_1,\dots,i\omega_{2k}
\rbrace$, with $\omega_i\in \R\setminus\{0\}$ for $i=1,\dots,2k$.
Let us for the context of an averaging analysis assume that the  
vector fields on the r.h.s. of (~\ref{parttzZeq}) are real 
analytic with respect to $(y,z)$.
We apply a complex linear coordinate transformation that
diagonalizes $\Al$, and denote the complexified,
new coordinates and vector fields again by $(y,z)$, and $Y(y,z)$, $Z(y,z)$, 
respectively, by which we find
\eqn\label{beginning}
		\partial_t ( y_t,z_t) =  \big({\rm diag}(i\omega) y_t 
         + Y(y_t,z_t) \,,\,Z(y_t,z_t)\big) \; \in \C^{2k}\times\C^{2(n-k)}\;, \eeqn
where $\omega:=(\omega_1,\dots,\omega_{2k})$.
Complexifying (~\ref{parttzZeq}), the continuation of $\cg$ into 
$\C^{2n}$ is defined as the common zeros of $Y(0,z)$ and $Z(0,z)$ for 
$z\in\C^{2(n-k)}$.

We next introduce polar coordinates $(I,\phi)\in\R^{2k}\times[0,2\pi)^{2k}$ and 
$(J,\theta)\in\R^{2(n-k)}\times[0,2\pi)^{2(n-k)}$ in terms of
$y^r =: e^{i\phi_r} I^r$ and $z^s=: e^{i\theta_s} J^s$,
with $r=1,\dots,2k$ and $s=1,\dots,2n-2k$. In particular,
$I\in\R^{2k}$, $J\in\R^{2n-2k}$,
$\phi\in[0,2\pi]^{2k}=\TT^{2k}$ (the $2k$-dimensional torus), and
$\theta\in[0,2\pi]^{2n-2k}=\TT^{2n-2k}$. For brevity, let 
$e^{i\phi} v:=(e^{i\phi_1}v^1,\dots,e^{i\phi_{2k}}v^{2k})$ and
$e^{i\theta}w :=(e^{i\theta_{1}} w^{1},\dots,e^{i\theta_{2(n-k)}} w^{2(n-k)})$, for
$v\in\C^{2k}$ and
$w\in\C^{2n-2k}$.
(~\ref{beginning}) is then easily seen to 
be equivalent to (the dot abbreviates $\partial_t$)
\eqn\dot{I}  =  {\rm Re}\lbrace e^{-i\phi}
           Y(e^{i\phi}I,e^{i\theta}J)\rbrace&, &
       \dot{\phi} = \omega + {\rm Im}\lbrace e^{-i\phi}
       {\rm diag}(\partial_I) Y(e^{i\phi}I,e^{i\theta}J)\rbrace . 
		\label{castor}\\
		\dot{J} =  {\rm Re}\lbrace e^{-i\theta}
           Z(e^{i\phi}I,e^{i\theta}J)\rbrace & , &
       \dot{\theta} = {\rm Im}\lbrace e^{-i\theta}
       {\rm diag}(\partial_J) Z(e^{i\phi}I,e^{i\theta}J)\rbrace .
		\label{pollux} 
\eeqn
Let us assume that $\epsilon:=|I(0)|\ll 1$, and  $|J(0)|\leq
O(\epsilon^2)$. We then introduce rescaled variables
$I\rightarrow\epsilon I$ and $J\rightarrow \epsilon^2 J$.

Analyticity of $Y(y,z)$ and $Z(y,z)$ with respect to $(y,z)$ implies that the
power series expansion with respect to $e^{i\phi}I$ and $e^{i\theta}J$ 
converges for $\epsilon$ sufficiently small.
Accordingly, (~\ref{castor}) and (~\ref{pollux}) yield
\eqn \label{bog1}
     \dot{I}^r&=&\sum_{|m|+|p|\geq 2}\epsilon^{|m|+2|p|-1}F^r_{mp}(I,J)
       e^{i(\langle m,\phi\rangle-\phi_r)}
      e^{i\langle p,\theta\rangle}\\ \label{bog2}
     \dot{J}^s&=&\sum_{|m|+|p|\geq 2}\epsilon^{|m|+2|p|-2}G^s_{mp}(I,J)
       e^{i\langle m,\phi\rangle}e^{i\langle p,\theta\rangle} \\ \label{bog3}
     \dot{\phi}_r&=&\omega_r +
       \sum_{|m|+|p|\geq 2}\epsilon^{|m|+2|p|-1}\Phi_{r;mp}(I,J)
       e^{i(\langle m,\phi\rangle-\phi_r)}
       e^{i\langle p,\theta\rangle}\\ \label{bog4}
     \dot{\theta}_s&=&
       \sum_{|m|+|p|\geq 2}\epsilon^{|m|+2|p|-2}\Theta_{s;mp}(I,J)
       e^{i\langle m,\phi\rangle}e^{i\langle p,\theta\rangle} ,\eeqn
where we introduced the multiindices $m\in\Z^{2k}$ and $p\in\Z^{2n-2k}$, with
$|m|:=\sum|m_r|$ and $|p|:=\sum|p_s|$. In this Fourier expansion with respect to
the $2\pi$-periodic angular variables
$\phi$ and $\theta$, every Fourier coefficient labeled by a pair of
indices $(m,p)$, is a homogenous polynomial of degree $|m|$ in  $I$,
and of degree $|p|$ in $J$.

If the components of $\omega$ are all mutually rationally independent, one may
consider the averaged quantities $f_t(\phi)\rightarrow\bar{f}_t:=
(2\pi)^{-n}\int_{\TT^{n}}d\phi f_t(\phi)$.
From (~\ref{beginning}), $Y(y,z)$ and $Z(y,z)$ are  
$O(|y|)$, thus their power series involve terms
$e^{i\langle m,\phi\rangle}$ with $|m|\geq 1$, but none with $|m|=0$. 
Averaging  
(~\ref{bog1}) $\sim$ (~\ref{bog4}) with respect to $\phi$ thus gives
\eqn \dot{\bar{I}}=\epsilon^2 \tilde{F}(\bar{I},\bar{J},\bar{\theta})
      \; \; , \; \; 
		\dot{\bar{J}}=0
		\; \; , \; \;  \dot{\bar{\theta}}=0 \eeqn
for some function $\tilde{F}$, where the bars account for averaged variables.
Thus, if we in addition assume that there exists a local degenerate almost
Lyapunov function with respect to $\cg\cap U$, it follows for the averaged
equations of motion that $|\bar I|$ is bounded for all $t$.
In particular, if the incommensurability condition holds, $|\bar J|$ is
then also bounded
for all $t$, and $a$ (respectively $0$) is, for the averaged system, stable.  
Based on these insights, and on intuition stemming from KAM and
Nekhoroshev theory, it is thus natural to 
conjecture the following stability criterion.

\begin{conj}\label{stabcriteriaconj}
Let $\crit_i\subset\cg$ be a connectivity component of the critical manifold,
and let $\xa\in\crit_i$, with 
{\rm spec}$\{DX_H^V(a)\}\setminus\{0\}=\lbrace i\omega_1,\dots,
i\omega_{2k}\rbrace$, and $\omega_i\in\R\setminus\lbrace0\rbrace$ for $i=1,\dots,2k$. 
Assume that (1)
the frequencies $\omega_r$ are rationally independent, and (2) that there 
exists a local degenerate, almost
Lyapunov function with respect to $\crit_i\cap U(a)$, in the sense of 
definition {~\ref{LDLFdef}}.
Then, $\xa$ is stable in the sense of Nekhoroshev.
In particular, condition (2) is always satisfied (by the 
Hamiltonian $H$) if the index
of $\crit_i$ is $\mu(\crit_i)=0$.
\end{conj}

\subsection{The Relationship to Sub-Riemannan Geometry}

To elucidate the geometric nature of the requirement of rationally
independent frequencies, we will now approach the discussion of critical 
stability from a different point of view. This discussion involves
issues that are central to sub-Riemannian geometry, \cite{BeRi,Ge,Gr2,Str}.

We study the time evolution map in a tubular $\epsilon$-vicinity of $U(a)\cap\cg$
by invoking a geometrically invariant Lie series that is adapted to the elements 
of the flag of $V$. By an asymptotic analysis, we explain the mechanism by which 
an instability can arise. The reason is that if the eigenfrequencies of the linear 
problem are not incommensurable, the problem of small divisors appears. 
This picture seems to be familiar from the perturbation theory of integrable
Hamiltonian systems, but we note once more that the lack of integrability here
originates from the non-holonomy of the constraints. A rigorous treatment of this last part
of the analysis is beyond our current scope, and left for future work.

\subsubsection{Dynamics Along the Flag of V}

Let $U$ denote a small open
neighborhood $U$  of $\xa\in\crit$, and assume
that $\cg:=\crit\cap U$ satisfies the genericity
condition of theorem {~\ref{Sardthm}}.

\begin{lem}\label{tubneigh}
Let  $\cg=\crit\cap U$ have the genericity property
formulated in theorem {~\ref{Sardthm}}.
Then, there exists $\epsilon>0$ such that
every point $x\in U$ with $d_R(x,\cg)<\epsilon$
is given by
$$x=\exp_s Y (\xa)\;\;\;\;,\;\;\;\;|s| < \epsilon$$
for some $Y\in\Gamma(V)$ with $\|Y\|_{g_M}\leq 1$,
$\xa\in\cg$
($\exp_s Y$ denotes the 1-parameter group of diffeomorphisms
generated by $Y$, with $\exp_0 Y=$id).
\end{lem}

\begin{proof}
We choose a spanning family $\lbrace Y_i\in\Gamma(V)\rbrace_1^{2k}$ of $V$,
with $\|Y_i\|_{g_M}= 1$.
If for all $\xa\in\cg$, $T_{\xa}\cg$ contains no subspace of $V_{\xa}$, then
$$\exp_1(t_1Y_1+\dots+t_{2k}Y_{2k})(\cg)\;\;\cap\;\;U$$
is an open tubular neighborhood of $\cg$ in $U$,
for $t_i\in (-\epsilon,\epsilon)$.
Because the normal space $N_{\xa}\cg$ is dual to the span of
the 1-forms $dF_i$ at $\xa$,
this condition is satisfied if and only if the matrix
$[dF_j(Y_i)]=[Y_i(Y_j(H))]$
is invertible everywhere on $\cg$.
According to proposition {~\ref{invert}}, this condition
is indeed fulfilled. \end{proof}

Hence, there is an element
$Y\in\Gamma(V)$ with $\|Y\|_{g_M}\leq 1$,
so that
$x=\Psi_\epsilon(\xa)$
for some $0<\epsilon\ll 1$.
Since $\xa\in\cg$, it is clear that under the flow
generated by $X_H^V$, $\tPhi_{\pm t}(\xa)=\xa$, thus
the solution of (~\ref{eqsofmo}) with initial condition $x$
is given by
\eqnn \Psi_\epsilon^t(\xa)\;:=\;
      \tPhi_t\circ\Psi_\epsilon(\xa)\;=\;\left(\tPhi_t\circ
      \Psi_{\epsilon}
      \circ\tPhi_{-t}\right)(\xa)\;.\eeqnn
$\Psi_\epsilon^t$ is, in particular, the 1-parameter group of diffeomorphisms
with respect to the variable $\epsilon$ that
is generated by the pushforward vector field
\eqn\label{pushf}Y_t(x)\;:=\;\tPhi_{t\;*}\;Y(x)\;=\;
      d\tPhi_{t}\circ Y(\tPhi_{-t}(x))\;,\eeqn
where $d\tPhi_t$
denotes the tangent map associated to $\tPhi_t$. From the group property $Y_{s+t}=\tPhi_{s\,*} Y_t$ follows that
\eqn\label{Ytder}\partial_t Y_t\;=\;\left.\partial_s\right|_{s=0}
      \tPhi_{s\,*} Y_t \;=\;[X_H^V,Y_t]\;,\eeqn
everywhere in $U$.

Next, we pick a local spanning
family  $\left\lbrace Y_i\in\Gamma(V)\right\rbrace_{i=1}^{2k}$
for $V$ that  satisfies
$\omega(Y_i,Y_j)=\tJ_{ij}$,
with
$\tJ:=\left(\begin{array}{cc}0&\1_k\\-\1_k&0\end{array}\right)$.
Furthermore, defining $\theta_i(\cdot):=\omega(Y_i,\cdot)$,
$\PV=\tJ^{ij}Y_i\otimes \theta_j$,
where $\tJ^{ij}$ are the components of $\tJ^{-1}=-\tJ$. In particular,
\eqnn X_H^V\;=\;\PV(X_H^V)\;=\;-\;Y_i(H)\;\tJ^{ij}\; Y_j\;\eeqnn
in the basis $\left\lbrace Y_i\right\rbrace_{i=1}^{2k}$. 

The following proposition characterizes the orbit emanating from $x$
in terms of nested commutators with respect to $Y_t$.

\begin{prop}
Let $f,F_i\in C^\infty(U)$, where $F_i := Y_i(H)$, $i=1,\dots,2k$,
and assume that $F_i(\Psi_\epsilon^t(\xa))$,
$f(\Psi_\epsilon^t(\xa))$ are real analytic in $\epsilon$.
For $X,Y\in \Gamma(TM)$,
let $$\cL_Y^r X \;=\;[Y,\dots,[Y,X]]$$ denote the $r$-fold iterated
Lie derivative. Then, for sufficiently small $\epsilon$,
\eqn\label{dertflag}\partial_t f(\Psi_{\epsilon}^t(\xa))=-
       F_i(\Psi_\epsilon^t(\xa))\tJ^{ik}\sum_{r\geq 0}
      \frac{\epsilon^r}{r!} 
      (\cL_{Y_t}^{r} Y_k)(f\circ\Psi_\epsilon^t)(\xa)\;.\eeqn
\end{prop}

\begin{proof}
Clearly,
\eqn\partial_t f(\Psi_\epsilon^t(\xa))&=&X_H^V(f)
     (\Psi_\epsilon^t(\xa))\nonumber\\
     &=&-
     F_i(\Psi_\epsilon^t(\xa))\tJ^{ik}
     Y_k(f)(\Psi_\epsilon^t(\xa))\nonumber\\
     &=&-
     F_i(\Psi_\epsilon^t(\xa))\tJ^{ik}
     \left(\Psi_{\epsilon\,*}^t Y_k \right)
     (f\circ\Psi_\epsilon^t)(\xa)\;.\eeqn
Using the Lie series
$\Psi_{\epsilon\;*}^t  Y_k = \sum_r \frac{\epsilon^r}{r!} 
    \cL_{Y_t}^{\;r} Y_k$,
we arrive at the assertion. \end{proof}

\begin{prop}\label{commflag}
Assume that $Y_{t=0}\in\Gamma(V)$, and let $\lbrace Y_j\rbrace_1^{2k}$
be the given local spanning family of $V$.
Then, $\cL_{Y_t}^{\;i}Y_j\in\Gamma(V_i)$, where $V_i$ is the
$i$-th flag element of $V$.
\end{prop}

\begin{proof}
Since $\tPhi_{t\;*}:\Gamma(V)\rightarrow\Gamma(V)$, $Y_t$
is a section of $V$ for all $t$ if it is for $t=0$.
The claim immediately
follows from the
definition of the flag of $V$.
\end{proof}

Proposition {~\ref{commflag}} implies that there are functions
$a^i(t,\cdot)\in C^\infty(U)$, $i=1,\dots,2k$, so that
$Y_t(x) = a^i(t,x)Y_i$. The next proposition determines their
time evolution.

\begin{prop}
Let $Y_{t=0}=a_0^i Y_i$ define the initial condition, and introduce
the matrix
$$\Omega_x:=[Y_l(F_i)(x)\tJ^{ij}]\;.$$
Then, pointwise in $x$,
\eqn\label{abODE} a^m(t,x)&=&\left(\exp(-t\Omega_x)\right)^m_j a_0^j
      +F_j(x)R^{jm}_i(t,x)a_0^i \;,\eeqn
where
$$R^{jm}_i(t,x):= \tJ^{jl}\tJ^{nk} \int_0^t ds
     (\exp(-(t-s)\Omega_x))^m_k
      \omega\left([Y_l,\tPhi_{s\,*}Y_i], Y_n\right)\;.$$
\end{prop}

\begin{proof}
The initial condition at $t=0$ is given by
$Y_0=a^i_0 Y_i$, that is, by $a^i(0,x)=a^i_0$.
Thus, by the definition of $Y_t$ in  (~\ref{pushf}),
one has $Y_t=a^i_0\, \tPhi_{t\;*}Y_i$,
so that
$a^i(t,x)Y_i=a^i_0\tPhi_{t\;*}Y_i$.
From $\omega(Y_i,Y_j)=\tJ_{ij}$, $\tJ_{ik}=-\tJ_{ki}$ and
$\tJ_{im}\tJ^{ml}=-\delta^l_i$,
$$
		a^l(t,x)=-a^i_0\omega\left(\tPhi_{t\;*}Y_i\,,\,Y_{j}\right)
     	\tJ^{jl}\;.$$
Now, taking the $t$-derivative on both sides of the equality sign,
one finds
\eqnn\partial_t a^m(t,x)&=&- a^i_0
     \omega\left([X_H^V,\tPhi_{t\;*}Y_i]\,,\,Y_{k}\right)\tJ^{km}\\
     &=&- a^i(t,x) Y_i(F_j)(x) \tJ^{jm} \nonumber\\
     &&
     - a^i_0 F_j(x) \tJ^{jl} \tJ^{km} 
      \omega\Big([Y_l,\tPhi_{t\;*}Y_i]\,,\,Y_{k}\Big)\;.\eeqnn
Using the variation of constants formula pointwise in $x$, one
arrives at the assertion.
\end{proof}

\subsubsection{Non-holonomy and small divisors}\label{Loptsubsubsect}
Using the description of the dynamics in the vicinity of $a$ derived above,
we will here use the small parameter $\epsilon$
for an asymptotic expansion. The intention of this part of the discussion, 
which is not rigorous, is to explain the geometric origin of the incommensurability 
condition on frequencies in conjecture {~\ref{stabcriteriaconj}}.

We consider the following simplified situation:
\newcounter{nn0}
\begin{list}
  {(\arabic{nn0})}{\usecounter{nn0}\setlength{\rightmargin}{\leftmargin}}
\item $\Omega_x=\Omega$, constant for all $x$ in $U$.
\item spec$\{\Omega\}=\lbrace i\omega_1,\dots,i\omega_{2k}\rbrace$, with
   $\omega_r\in \R$.
\item $\|\Omega\|:=\sup_r|\omega_r|\ll\frac{1}{\epsilon}$.
\end{list}

Let us briefly comment on the generic properties
of $\lbrace\omega_r\rbrace$.
Writing $\Omega=\tJ A$,
we decompose the matrix $A=[Y_i(Y_j(H))(\xa)]$
into its  symmetric and antisymmetric parts
$A_+$ and $A_-$, respectively.
$A_-=[[Y_j,Y_i](H)(\xa)]/2$ vanishes
if $V$ is integrable, which one deduces from 
$X_H|_{\xa}\in V_{\xa}^\perp$ for all $\xa\in \cg$, and the Frobenius condition.
The linear system $\underline{\dot{a}}=\tJ A_+ \underline{a}$ 
is Hamiltonian, hence the spectrum of $\tJ A_+$ consists of complex 
conjugate pairs of eigenvalues in $i\R$ if it is purely imaginary
(here, $\underline{a}:=(a^1,\dots,a^{2k})$).
Considering $\tJ A_-$ as a perturbation of $\tJ A_+$, we may generically 
assume that all frequencies $\omega_r$ are distinct from one another,
and that there are both negative and positive frequencies.

Under the simplifying assumptions at hand, let us compute (~\ref{abODE}) to
leading order in $\epsilon$.
From (~\ref{abODE}), one infers
$$Y_t=a^j_0(\exp(-t\Omega))^i_j  Y_i + \sum_i  O(|x|) Y_i\;,$$
since $|F_j(x)|=O(|x|)=O(\epsilon)$, which follows from $F_j(\xa)=0$.
Thus,
$$
		[Y_t,X]= a^j_0 \exp(-t \Omega)_j^i [Y_i,X] +
      	\sum_i O(\epsilon) [Y_i,X]
     	+\sum_i O(1) Y_i 
$$
for all $X\in\Gamma(TM)$, and $x\in U_\epsilon(\xa)$.
Assuming that all objects in question are $C^\infty$,
iterating the Lie bracket $\cL_{Y_t}$ $r$ times produces
$$
		\Big(\prod_{m=1}^r
       a^{j_m}_0 \big(\exp(-t \Omega)\big)^{i_m}_{j_m} + O(\epsilon)
       \Big)
       [Y_{i_1},[Y_{i_2},\dots,[Y_{i_r},Y_l]\cdots]]\;,
$$
plus a series of terms with less than  $r$ nested Lie commutators
that contribute to higher order corrections.

Let us, for the discussion of the leading order terms along each
flag element of $V$, omit the relative errors of order $O(\epsilon)$. 
By the assumption of smoothness, our considerations are valid for
$t\leq O(\epsilon^{-1})$. Let us consider the term
\eqn\label{Vrcontr}
		F_i(\Psi_\epsilon^t(\xa)) 
      	\tJ^{ik} \left(\cL^{r}_{Y_t}Y_k\right) (f\circ \Psi_\epsilon^t)(\xa)\;,
\eeqn
for fixed $r$. It is easy to see that
\eqn\label{Fiser} F_i(\Psi_\epsilon^t(\xa)) = Y_t(F_i)(\xa) +
      	O(\epsilon^2)\;,\eeqn
due to $F_i(\xa)=0$.
Therefore,
\eqn F_i(\Psi_\epsilon^t(\xa))\tJ^{ik}=\epsilon
     \exp\left(-t\Omega\right)^m_j a_0^j\Omega^k_m+
     O(\epsilon^2)\;,\eeqn
from a straightforward calculation.

Hence, the
terms with $r$ nested commutators in (~\ref{Vrcontr}) are
\eqnn\frac{\epsilon^{r+1}}{r!}  a^i_0 
       \exp(-t\,\Omega)^j_i \Omega_j^l\,\Big(\prod_{m=1}^r
       a^{j_m}_0\, \left(\exp(-t\,\Omega)\right)^{i_m}_{j_m}
       \Big)
       [Y_{i_1},[ \dots,[Y_{i_r},Y_l]\cdots]](f)(\xa)\\
       +  O(\epsilon^{r+2})\;,\eeqnn
as long as $dist_R(\Psi_\epsilon^t(\xa),\xa)\leq O(\epsilon)$.
This implies that for $f\in C^\infty(U)$,
\eqn f(\Psi_\epsilon^t(\xa)) &\sim& f(\xa) + \sum_{r\geq 0} 
       \frac{\epsilon^{r+1}}{r!} \int_0^t ds  a^i_0 
       \exp(-s \Omega)^j_i \Omega_j^l 
       \nonumber\\
       && 
       \Big(\prod_{m=1}^r
       a^{j_m}_0 \left(\exp(-s\;\Omega)\right)^{i_m}_{j_m}
       \Big)
       [Y_{i_1},\dots,[Y_{i_r},Y_l]\cdots](f)(\xa)\;,
       \label{fflagexp}\eeqn
up to relative errors of higher order in $\epsilon$ for every fixed $r$.

If $f$ is chosen as the $i$-th coordinate function $x^i$, so that
$f(\Psi_\epsilon^t(\xa))=x^i_t$,
the quantity $[Y_{i_1},\dots,[Y_{i_r},Y_l]\cdots](f)(\xa)$ is the
$i$-th coordinate of the vector field defined by the brackets at $\xa$.
Consequently, (~\ref{fflagexp}) is the component decomposition of
$x^i_t$ relative to the flag of $V$ at $\xa$, to leading order in
$\epsilon$.

By the given simplifying assumptions,
spec$\{\Omega\}\subset i\R\setminus\{0\}$, and the norm of 
$\exp(-s\Omega)$ is 1, independently of $s$. Consequently, the integrand
of (~\ref{fflagexp}) is bounded for all $s$. It follows that
if the $r$-th integral in the sum diverges, it will become apparent only for
$t\geq O\big(\frac{1}{\epsilon^{r}}\big)$. This would correspond to an
instability along the direction of the flag element $V_r$. 
While the leading term with $r=0$ is bounded for all $t$,   
terms with $r>0$ can diverge.

We next write
\eqn\underline{a}(s)=\exp(-s \Omega) \underline{a}_0
     = \sum_{\alpha=1}^{2k} 
     A_\alpha \underline{e}_\alpha \exp(-i\omega_\alpha s)
     \;,\eeqn
where $\lbrace \underline{e}_\alpha\rbrace$ is an orthonormal
eigenbasis of $\Omega$ with respect to the standard
scalar product in $\C^{2k}$, and spec$\{\Omega\}=\lbrace i\omega_\alpha\rbrace$.
The amplitudes $A_\alpha\in \C$ are determined by the initial condition
$a^i(t=0)=a_0^i$, which we assume to be nonzero.
By linear recombination of the vector fields $Y_i$, one can set
$e^i_\alpha=\delta_{i,\alpha}$. Then,
(~\ref{fflagexp}) can be written as
\eqn\label{intnestcomm}\sum_{r\geq 0} \frac{\epsilon^{r+1}}{r!} 
       \sum_{l;i_1,\dots,i_r} I_{l;i_1,\dots,i_r}(t) 
        [Y_{i_1},\dots,[Y_{i_r},Y_l]\cdots](f)(\xa)\;,\eeqn
where
\eqn\label{Iindt}  I_{l;i_1,\dots,i_r}(t)&:=&
       \int_0^t ds \omega_l A_l  \Big(\prod_{m=1}^r
       A_{j_m} \Big) e^{-is (\omega_l + \sum_{m=1}^r
       \omega_{j_m} )}\nonumber\\
       &=&\frac{i\omega_l A_l }{\omega_l + \sum_{m=1}^r
       \omega_{j_m}} \Big(\prod_{m=1}^r
       A_{j_m} \Big)   \Big( 
       e^{ -it (\omega_l + \sum_{m=1}^r
       \omega_{j_m} )} - 1 \Big)\;.
       \label{Ili1dotsirdef}\eeqn
We note that  the sum of frequencies (of generically indefinite
signs) in the denominator on the last line
raises the problem of small divisors. We also remark that evidently, 
the nested commutators vanish if all indices
$i_1,\dots,i_r,l$ have equal values, as it should be (otherwise,
the solutions would always diverge).

\subsubsection{Rational frequency dependence and blow-up of solutions}

Let us next discuss the situation in which the small divisors approach zero.
To this end, we introduce the set
\eqn\I^{(r)}(t)\;:=\;\left\lbrace I_{l;i_1,\dots,i_r}(t)
      \right\rbrace_{l,i_j=1}^{2k}\;
      \setminus\; \left\lbrace I_{l;l,\dots,l}(t)
      \right\rbrace_{l=1}^{2k}\;,\eeqn
which we endow with the norm
$\|\I^{(r)}(t)\|:=\sup_{I(t)\in\I^{(r)}(t)} |I(t)|$, and let
$\|A\|:=\sup_{i=1,\dots,2k}\lbrace|A_i|\rbrace$,
where $A_i$ are $\C$-valued amplitudes.

Furthermore, let
$\Aa:=\lbrace\omega_1,\dots,\omega_{2k}\rbrace$,
and let 
\eqn\Aa_r := \underbrace{\Aa + \cdots + 
    \Aa}_{r\;{\rm times}} \;,\eeqn
denote its $r$-fold sumset,
which is the set containing all sums of $r$ elements of $\Aa$.

For two sets of real numbers $\Aa$ and ${\mathfrak B}$, we define  
\eqn d(\Aa,{\mathfrak B}) := \inf_{i,j}\Big\lbrace  |a_i-b_j |\;\Big|\;
     a_i \in \Aa\;,\;b_j \in {\mathfrak B} \Big\rbrace \;.
\eeqn
Then, it follows from (~\ref{Ili1dotsirdef}) that if $d(\Aa_r,-\Aa)>0$,
\eqn\|\I^{(r)}(t)\| \leq d(\Aa_r,-\Aa)^{-1} \|\Omega\|
      \|A\|^r\;\eeqn
(the sum over frequencies $\sum_{m=1}^r
\omega_{j_m}$ in (~\ref{Ili1dotsirdef}) is
an element of $\Aa_r$, and can only equal $-\omega_l$ if
$d(\Aa_r,-\Aa)=0$).
However, if $d(\Aa_r,-\Aa)=0$, there is a tuple of indices
$\lbrace l;i_1,\dots,i_r\rbrace$ such that
\eqn I_{l;i_1,\dots,i_r}(t) = - t \omega_l A_l \prod_{m=1}^r
       A_{j_m} \;,\eeqn
in case of which $\|\I^{(r)}(t)\| \sim t$,
that is, a divergence linear in $t$ for large $t$
(recalling that the present asymptotic considerations require $t\leq\epsilon^{-1}$).
Only if there are simultaneously positive and negative frequencies,
$d(\Aa_r,-\Aa)=0$ is possible, but due to the remark
at the beginning of subsection {~\ref{Loptsubsubsect}},
this situation must generically assumed to be given.

As an illustration, the following picture holds for $r\leq2$.
The fact that for $r=0$, $\|\I^{(0)}(t)\|$ is bounded for all $t$ is clear.
For $r=1$, the first flag element $V_1=[V,V]$ is in question. The condition
for the emergence of a divergence is that $d(\Aa,-\Aa)=0$.
This is precisely given if there is a pair of frequencies $\pm\omega_i$
of equal modulus, but opposite sign. For $r=2$, assuming that
$d(\Aa,-\Aa)>0$, the condition $d(\Aa_2,-\Aa)=0$ implies that there
is a triple of frequencies such that $\omega_{i_1}+\omega_{i_2}=
-\omega_{i_3}$, $i_j\in\lbrace1,\dots,2k\rbrace$. If this occurs,
the solution will diverge in the direction of the second
flag element, $V_2=[V,[V,V]]$. The discussion for $r>2$ continues in
the same manner.

Hence, our conclusion from this asymptotic analysis
is that if $d(\Aa_r,-\Aa)=0$ for some $r$, then 
$\|\I^{(r)}(t)\|= O( t)$ for $t\rightarrow\infty$.

The physical insight gained from the above discussion can be summarized
as follows. If the frequencies of the linearized
problem fail to satisfy the incommensurability condition
$d(\Aa_r,-\Aa)>0$ for all $r$, the equilibrium $\xa$ is unstable.
However, the time required for an orbit to exit from a Riemannian
$\epsilon$-neighborhood $U_\epsilon(\xa)$ is very large.
In fact, assuming that $d(\Aa_r,-\Aa)=0$ for some $r\leq \rV $
(the degree of non-holonomy of $V$), a
time $t\sim O(\frac{1}{\epsilon^{r}})$ is necessary to exit
from  $U_\epsilon(\xa)$ in the direction of the flag element $V_r$ 
(due to the factor $\frac{\epsilon^r}{r!}$ in (~\ref{fflagexp})).
We note that the orbit does not drift out from $U_\epsilon(\xa)\cap\cg$
in the direction of $V_{\xa}$ owing to the existence of a local degenerate
Lyapunov function required in the conjectured stability criterion. 
Therefore, this discussion suggests that the incommensurability
condition imposed on the frequencies of the linearized system
can indeed not be omitted.

\subsubsection{Instabilities in the Context of Carnot-Caratheodory Geometry}

The 
constrained Hamiltonian system
$(M,\omega,H,V)$ shares many characteristics with systems typically encountered
in sub-Riemannian geometry \cite{BeRi,Ge,Gr2,Str}.
The natural metric structure in this context is given by
the Carnot-Caratheodory distance function
$dist_{C-C}$  induced by the Riemannian metric $g$.
It assigns to a pair of points
$x,y\in M$ the length of the shortest $V$-horizontal $g$-geodesic.

If $V$ satisfies the Chow condition, $dist_{C-C}(x,y)$ is
finite for all $x,y\in M$, by the Rashevsky-Chow theorem \cite{BeRi,Gr2}.
In this case, the Carnot-Caratheodory $\epsilon$-ball
$$B^{C-C}_\epsilon(\xa) := \Big\lbrace x\in M \Big| 
        dist_{C-C}(x,\xa) < \epsilon \Big\rbrace$$
is open in $M$.

If $V$ fails to satisfy the Chow condition, pairs of points
that cannot be joined by $V$-horizontal
$g_M$-geodesics are assigned a Carnot-Caratheodory distance $\infty$.
Then, $M$ is locally foliated into
submanifolds $N_\lambda$
of dimension $(2n-{\rm rank}V_{\rV })$ 
(we recall that $\rV $ denotes the degree of non-holonomy of $V$),
with $\lambda$ in some index set,
which are integral manifolds of the
(necessarily integrable)
final element $V_{\rV }$ of the flag of $V$.
On every $N_\lambda$, the distribution
$V_\lambda:=j_\lambda^* V$ satisfies the Chow condition,
where $j_\lambda:N_\lambda\rightarrow M$ is the inclusion.
Therefore, all points $x,y\in N_\lambda$ have a finite
distance with respect to the Carnot-Caratheodory metric induced by the
Riemannian metric  $j_\lambda^*g_M$.
Every leaf $N_\lambda$
is an invariant manifold of the flow $\tPhi_t$.

Let $\lbrace Y_{i_r}\rbrace_{r=1}^{\rV }$ denote a local spanning family 
of $TM$ such that $\lbrace Y_{i_r}\rbrace$ spans the flag element
$V_r$. Let the $g$-length of all $Y_{i_r}$'s be 1.
Then, we define the 'quenched' box
$${\rm Box}_\epsilon(x) :=  \Big\lbrace
      \exp_1\Big(\sum_{r=1}^{\rV }
      \epsilon^r\sum_{i_r=1}^{{\rm dim}V_r}t_{i_r}\;Y_{i_r}\Big)(x)\;
      \Big|\;t_{i_r} \in (-1,1)\Big\rbrace\;$$
in $N_\lambda$, where $\lambda$ is suitably picked so that $x\in N_\lambda$.
Evidently, if $V$ satisfies Chow's condition, $N_\lambda=M$.
According to the ball-box theorem \cite{BeRi,Gr2}, there are constants
$C>c>0$, such that
$${\rm Box}_{c\epsilon}(x)\;\subset\;B^{C-C}_\epsilon(x)\;
      \subset\;{\rm Box}_{C\epsilon}(x)\;.$$
Carnot-Caratheodory $\epsilon$-balls can be approximated
by quenched boxes in Riemannian geometry.

The above perturbative results imply that
if there is some $r<\rV $, for which
$d(\Aa_r,-\Aa)=0$,
the flow $\tPhi_t$ blows up the quenched boxes,
and thus the Carnot-Caratheodory $\epsilon$-ball
around $\xa\in\cg$, linearly in $t$, and along the direction of $V_r$.
In fact,
$B^{C-C}_\epsilon(\xa)$ is widened
along $V_r$ at a rate linear in $t$. For
$t=O(\frac{1}{\epsilon})$, $\tPhi_t$ maps the
Carnot-Caratheodory $\epsilon$-ball containing the initial condition
to a Carnot-Caratheodory ball of radius  
$O(1)$. Thus, in the context of Carnot-Caratheodory geometry, these
instabilities, which have no counterpart in systems
with integrable constraints, are far more significant 
than in the Riemannian picture.

\section{\nbf AUTONOMOUS NON-HOLONOMIC SYSTEMS IN CLASSICAL MECHANICS}
\label{sectionIV}

In this main section, we focus on the analysis of non-holonomic mechanical
systems, and their relationship to the constrained Hamiltonian
systems considered previously.
The discussion is restricted to  linear non-holonomic, {\em Pfaffian}
constraints.

Let $(Q,g,U)$ be a Hamiltonian mechanical system, where
$Q$ is a smooth Riemannian $n$-manifold with
a $C^\infty$ metric tensor $g$, and where
$U\in C^\infty(Q)$ denotes the potential energy.
No gyroscopic forces are taken into consideration.
Let $g^*$ denote the induced Riemannian metric on the cotangent bundle
$T^*Q$.
For $X\in\Gamma(TM)$, let $\theta_X$ be the 1-form defined by
$\theta_X(Y)=g(X,Y)$ for all  $Y\in\Gamma(TQ)$. Clearly,
$g(X,Y)=g^*(\theta_X,\theta_Y)$ for all  $X,Y\in\Gamma(TQ)$.

The K\"ahler metric of the previous discussion, also denoted
by $g$, will not appear in this section. From here on,
$g$ will denote the Riemannian metric on $Q$, which should not give
rise to any confusion.

In a local trivialization of
$T^*Q$, a point $x\in T^*Q$ is represented by a tuple $(q^i,p_j)$,
where $q^i$ are coordinates on $Q$,
and $p_k$ are
fibre coordinates in $T^*_q Q$, with $i,j=1,\dots,n$.
The natural symplectic 2-form associated to $T^*Q$, can be written in
coordinates as
$$\omega_0 = \sum_i dq^i \wedge dp_i\;=\;-d\theta_0\;. $$
$\theta_0=p_i dq^i$ is referred to as the symplectic 1-form.

We will only consider Hamiltonians of the form  
\eqn\label{hamham}H(q,p)\;=\;\frac{1}{2}\;g_q^*(p,p)\;+\;U(q)\; .
      \eeqn
In local bundle coordinates, the corresponding Hamiltonian
vector field $X_H$ is given by
$$X_H\;=\; \sum_i \left( (\partial_{p_i}H)\partial_{q^i} -
      (\partial_{q^i}H)\partial_{p_i}\right)\; .$$
The orbits of the associated Hamiltonian flow $\Phi_t$ satisfy
\eqn\label{herodot}\dot{q}^i = \partial_{p_i} H(q,p)\;\;\;\;,\;\;\;\;
         \dot{p}_j = -\partial_{q^j}H(q,p) \;.\eeqn
The superscript dot abbreviates $\partial_t$, and will be used throughout
the discussion.

Let ${\mathcal A}_I$ denote the space of
smooth curves $\gamma:I\subset\R\rightarrow T^*Q$, with $I$ compact
and connected, and let $t$ denote a coordinate on $\R$.
The basis one form $dt$ defines a measure on $\R$.
The action functional is defined by
${\mathcal I}:{\mathcal A}_I\rightarrow \R$,
\eqn\label{sagittarius}{\mathcal I}[\gamma]&=&\int_I\;dt\;
      \left(\gamma^*\theta_0\; -\;  H\circ\gamma \right)\\
      &=&\int_I \;dt\;\left(\sum
        p_i(t)\dot{q}^i(t)-H(q(t),p(t))\right)\;,\nonumber\eeqn
with
$\dot{\gamma}=\sum(\dot{q}^i\partial_{q^i}+\dot{p}_i\partial_{p_i})$.
Denoting the base point projection by
$\pi:T^*Q\longrightarrow Q$,
let
$c:=(\pi\circ\gamma):I\longrightarrow Q$
denote the projection of $\gamma$ to $Q$.
We assume that
$\|c(I)\|$ is sufficiently small so that
solutions of (~\ref{herodot}) exist, which connect the end points
$c(\partial I)$.
Among all curves
$\gamma:I\rightarrow T^*Q$ with fixed projected
endpoints $c(\partial I)$, the ones that extremize ${\mathcal I}$ are
physical orbits of the system.

\subsection{ Linear Non-Holonomic Constraints}

Let us next impose linear, 'Pfaffian'
constraints on the Hamiltonian mechanical system $(Q,g,U)$, 
by adding a rank $k$ distribution $W$
over $Q$ to the existing data, and  
invoke the H\"older variational principle, \cite{Ar1},
that generates the correct physical flow on $T^*Q$.
The orbits of the
resulting constrained dynamical system possess
$W$-horizontal projections to $Q$.
  
We introduce the $g$-symmetric projection tensor associated to $W$
given by
$\alph=\alph^2:TQ\rightarrow TQ$, with
$${\rm Ker}(\alph)\;=\;W^\perp\;\;\;\;,\;\;\;\;
       \alph(X)\;=\;X\;\;\;\;\forall\;X\;\in\;\Gamma(TQ)\;,$$
and its orthogonal complement $\bbeta=\1-\alph$. 
We note that in local coordinates, $\alph$ is represented by a 
$n\times n$ matrix of rank $k$.  The dual of $W$, denoted by $W^*$,
is defined as the image of $W$ under the
isomorphism $g:TQ\rightarrow T^*Q$, and likewise for
$(W^*)^\perp := g\circ W^*$. The corresponding $g^*$-orthogonal
projection tensors on $T^*Q$ are denoted by 
$\alph^\dagger$ and $\bbeta^\dagger$, respectively.
Our inspiration to introduce $\alph$ and $\bbeta$ for this
analysis stems from \cite{Bra}.
 
\subsubsection{ Dynamics of the Constrained Mechanical System}

Next, we derive the equations of motion of the constrained
mechanical system from the  H\"{o}lder  variational principle.
For a closely related approach to the Lagrangian theory of constrained
mechanical systems, cf.
\cite{CaFa}.

\begin{defn}
A projective $W$-horizontal curve in $T^*Q$
is an embedding $\gamma:I\subset\R\hookrightarrow T^*Q$
whose image
$c=\pi\circ\gamma$ under base point projection
$\pi:T^*Q\rightarrow Q$ is tangent to $W$.
\end{defn}

Let $\gamma_s:I\rightarrow T^*Q$, with $s\in[0,1]$, be a smooth
one parameter family of curves for which the end points
$c_s(\partial I)$ are
independent of $s$ (where $c_s:=\pi\circ\gamma_s$).

\begin{defn}
A $W$-horizontal variation of a projective $W$-horizontal curve $\gamma$
is a smooth one parameter family
$\gamma_s:\R\rightarrow T^*Q$, with $s\in[0,1]$, for which
$\frac{\partial}{\partial s}(\pi\circ\gamma_s)$ is tangent to $W$,
and $\gamma_0=\gamma$.
\end{defn}

Let
$$\delq^i(t):=\left.\partial_s\right|_{s=0}q^i(s,t)\;\;\;\;,\;\;\;\;
     \delp_k(t):=\left.\partial_s\right|_{s=0}p_k(s,t)\;.$$
To any $W$-horizontal variation $\gamma_s$ of a $W$-horizontal
curve $\gamma_0$ with fixed projections of the boundaries
\eqn\label{varcond}(\pi\circ\gamma_s)(\partial I)=
      (\pi\circ\gamma_0)(\partial I)\;,\eeqn
so that $\delq^i|_{\partial I}=0$, we associate
the action functional
$${\mathcal I}[\gamma_s]=\int_I \left(\sum
        p_i(s,t)\dot{q}^i(s,t)-H(q(s,t),p(s,t))\right)dt\;. $$

\begin{defn} (H\"older principle)
A physical orbit of the constrained mechanical system
$(Q,g,U,W)$ is a projective $W$-horizontal curve $\gamma_0:I\rightarrow T^*Q$
that extremizes ${\mathcal I}[\gamma_s]$ among all $W$-horizontal
variations $\gamma_s$ which satisfy (~\ref{varcond}).
\end{defn}

Hence,  if
\eqn\label{varact}
     \delta{\mathcal I}[\gamma_s]\;=\;\sum p_i \delq^i|_{\partial I} +\int_I
     \sum \left((\dot{p}_i-\partial_{q^i}H)\delq^i -
     (\dot{q}^i+ \partial_{p_i}H)\delp_i\right)\;=\;0\eeqn
for all $W$-horizontal variations of $\gamma_0$
that satisfy $\delq^i|_{\partial I}=0$,
then $\gamma_0$ is a physical orbit.

\begin{thm}\label{dae}
In the given local bundle chart,
the Euler-Lagrange equations of the H\"older variational
principle are the differential-algebraic relations
\eqn\label{qdotcon} \dot{q} &=& \alph(q)\partial_p H(q,p)
    \\ \label{pdotcon} \alph^\dagger(q)\dot{p} &=& -\alph^\dagger(q)
       \partial_q H(q,p) \\
       \label{physleaf} \bbeta(q)\partial_p H(q,p)&=&0\;.\eeqn
\end{thm}

\begin{proof}
The boundary term
vanishes due to $\delq^i|_{\partial I}=0$.

For any fixed value of $t$, one can write
$\delq(t)$ as
$$\delq(t) \;=\; \sum_{\alpha=1}^k \;f_\alpha(q(t))\; Y_\alpha(q(t))\; ,$$
where $Y_\alpha$ is a
$g$-orthonormal family of vector fields over $c(I)$ that spans $W_{c(I)}$.
Furthermore, $f_\alpha\in C^\infty(c(I))$ are
test functions obeying the boundary condition
$f_\alpha(c(\partial I))=0$.

Since $f_\alpha$ and
$\delp$ are  arbitrary, the terms in (~\ref{varact}) that
are contracted with $\delq$, and the ones that are contracted with
$\delp$ vanish independently.
In case of $\delq$, one finds
$$\int_I\; dt\;f_\alpha\;\;
      (\dot{p}+\partial_q H)_i\;Y_\alpha^i\;=0$$
for all test functions $f_\alpha$. Thus,
$(\dot{p}+\partial_q H)_i\,Y_\alpha^i=0$ for all
$\alpha=1,\dots,k$, or
equivalently, $\alph^\dagger(\dot{p}+\partial_q H)=0$,
which proves (~\ref{pdotcon}).

Since $\gamma_0$
is $W$-horizontal, $\bbeta(q)\dot{q}=0$,
so the $\delp$-dependent term in
$\delta{\mathcal I}[\gamma_s]$ gives
\eqnn \int_I dt\; (\dot{q}-\alph\partial_p H)^i\;(\alph^\dagger\delp)_i
      \;\;+\;\;\int_I dt\; (\bbeta\partial_p H)^i\;(\bbeta^\dagger\delp)_i
      \;=\;0\;.\eeqnn
The components of $\delp$ in the images of
$\alph^\dagger(q)$ and $\bbeta^\dagger(q)$ can be varied independently.
Thus, both terms on the second line must vanish separately,
as a consequence
of which one obtains (~\ref{qdotcon}) and (~\ref{physleaf}).
\end{proof}

\begin{defn}
The smooth submanifold 
$$
		\phys:=\Big\{(q,p)
    	\Big|\bbeta(q)\partial_p H(q,p) =0\Big\}
    	\subset T^*Q 
$$
locally characterized by (~\ref{physleaf}) is called the physical leaf.
\end{defn}

$\phys$ contains all physical orbits of the system, that is,
all smooth paths $\gamma:\R\rightarrow\phys\subset T^*Q$ that
satisfy the differential-algebraic relations of theorem {~\ref{dae}}.

\begin{thm}
Let $H$ be of the form (~\ref{hamham}). Then,
there exists a unique physical orbit $\gamma:\R^+\rightarrow\phys$ with
$\gamma(0)=x$ for every $x\in \phys$.
\end{thm}

\begin{proof}
We cover $\phys$ with local bundle charts of $T^*Q$ with coordinates $(q,p)$.
For the Hamiltonian (~\ref{hamham}), (~\ref{physleaf}) reduces to
$$\bbeta(q)\;g^{-1}(q)\;p\;=\;g^{-1}(q)\;\bbeta^\dagger(q)\;p\;=\;0\;,$$
where one uses the $g$-orthogonality of $\bbeta$.
Hence, (~\ref{physleaf}) is equivalent to  
$\bbeta^\dagger(q) p=0$. Since $\phys$ is the common zero level
set of the $n$ component functions $(\bbeta^\dagger(q) p)_i$,
every section
$$X\;=\;v^r(q,p)\;\partial_{q^r}\;+\;w_s(q,p)\;\partial_{p_s}$$
of $T\phys$ is annihilated by the 1-forms
$$d(\bbeta^\dagger p)_i\;=\;\partial_{q^r}(\bbeta^\dagger p)_i\;dq^r\;+\;
     \partial_{p_s}(\bbeta^\dagger p)_i \;dp_s$$
for $i=1,\dots,n$ (of which only
$n-k$ are linearly independent), on $\phys$.

This is expressed by 
\eqnn0&=& (v^r\partial_{q^r}) \bbeta^\dagger p\; +
    (w_s\partial_{p_s})\bbeta^\dagger p
     \\ &=&
    (v^r\partial_{q^r}) \bbeta^\dagger p\;+\;\bbeta^\dagger w\;,
\eeqnn
which shows that the components $v$ of $X$
determine the projection   
$\bbeta^\dagger w$. Hence, the components
$v$ and $\alph^\dagger w$ suffice  to uniquely reconstruct $X$.
Consequently, the right hand sides of (~\ref{qdotcon}) and
(~\ref{pdotcon}) determine a unique section $X$ of $T\phys$, so that every
curve $\gamma:\R^+\rightarrow\phys$, with arbitrary $\gamma(0)\in\phys$,
that satisfies
$\partial_t\gamma(t)=X(\gamma(t))$
automatically fulfills (~\ref{qdotcon}) $\sim$ (~\ref{physleaf}).
This proves the assertion.
\end{proof}

\subsubsection{ Equilibria}

The constrained Hamiltonian mechanical system
$(Q,g,U,W)$ possesses
\eqn\label{pegasus}\crit_Q :=\Big\lbrace q\in Q\Big|\alph^\dagger(q)
        \partial_q U(q)=0\Big\rbrace \eeqn
as its critical set.
An application of Sard's theorem fully analogous to the proof of
theorem {~\ref{Sardthm}} shows that generically,
this is a piecewise smooth, $n-k$-dimensional submanifold of $Q$,
(recall that the rank of
$\alph(q)$ is $k$).

\subsubsection{Symmetries} 
Let $G$ be a Lie group, and let
$\psi:G \rightarrow {\rm Diff}(Q)$, 
$h \mapsto \psi_h$ with $\Psi_e = {\rm id}$,
denote a group action. The constrained Hamiltonian mechanical system
$(Q,g,U,W)$ is said to exhibit a $G$-symmetry if the following hold.
(1) Invariance of the Riemannian metrics:
      $g\circ\psi_{h}=g$ and $g^*\circ\psi_{h}=g^*$ for all $h\in G$.
(2) Invariance of the potential energy:
      $U\circ\psi_{h}=U$ for all $h\in G$.
(3) Invariance of the distributions:
      $\psi_{h\,*}W=W$ and $\psi_h^* W^*=W^*$ for all $h\in G$.

\subsection{Construction of the Auxiliary Extension}

We are now prepared to embed the non-holonomic mechanical system into a
constrained Hamiltonian system of the type considered in the
previous sections.

To this end, we will introduce a set of generalized Dirac
constraints over the symplectic manifold $(T^*Q,\omega_0)$
in the way presented in section {~\ref{gendir}}.
They define a symplectic distribution $V$,
in a manner that the constrained Hamiltonian
system  $(T^*Q,\omega_0,H,V)$,
with $H$ given by (~\ref{hamham}), contains the constrained
mechanical system as a dynamical subsystem.
Thus, the auxiliary constrained Hamiltonian system 
$(T^*Q,\omega_0,H,V)$
extends the mechanical system in the sense announced in the
introduction.
An early inspiration for this construction stems
from \cite{SoBr}.
We require the following properties to be satisfied by $(T^*Q,\omega_0,H,V)$.

\newcounter{n2}
\begin{list}
{\roman{n2}.}{\usecounter{n2}\setlength{\rightmargin}{\leftmargin}}
\item $\phys$ is an invariant manifold under the flow $\tPhi_t$
      generated by (~\ref{eqsofmo}).
\item All orbits $\tPhi(x)$ with initial
      conditions $x\in\phys$  satisfy
      the Euler-Lagrange equations of the H\"older principle.
\item $\phys$ is marginally stable under $\tPhi_t$.
\item The critical set $\crit$ of $\tPhi_t$ is a vector
      bundle over $\crit_Q$, hence equilibria of the constrained mechanical
      system are obtained from equilibria of the extension by base point
      projection.
\item Symmetries of the constrained mechanical system extend
      to those of $\tPhi_t$.
\end{list}

Let us briefly comment on  (iii) $\sim$ (v).
(iii) is of importance for numerical simulations of the
mechanical system. (iv) makes it easy to extract
information about the behaviour of the
mechanical system from solutions of the auxiliary system.
Condition (v) allows to apply reduction theory to the
auxiliary system,
in order to reduce the constrained mechanical
system by a group action, if present.
The choice for $V$ is by no means unique,
and depending on the specific problem at hand,
other conditions than (iii) $\sim$ (v)
might be more useful.

\subsubsection{ Construction of $V$}

Guided by the above requirements,
we shall now construct $V$.

To this end, we
pick a smooth, $g^*$-orthonormal family of 1-forms
$\lbrace \zeta_I\rbrace_{I=1}^{n-k}$
with
$$\zeta_I\;=\;\zeta_{Ik}(q)\;dq^k\;,$$
so that locally,
$$\langle\lbrace\zeta_1,\dots,\zeta_{n-k}\rbrace\rangle =
      \left(W^*\right)^\perp\;.$$
The defining relationship $\bbeta^\dagger(q)p=0$ for $\phys$ is equivalent
to the condition
\eqn\label{holcon}f_I(q,p):= g_q^*(p,\zeta_I(q))=0 \;\;\;
      \forall I=1,\dots,n-k\;. \eeqn
It is clear that $f_I\in C^\infty(T^*Q)$.

\begin{enumerate}
\item  To satisfy conditions (i) and (iii), we require that the level surfaces
      \eqn\label{invman}
      \Mm_{\underline{\mu}}:=\left\lbrace(q,p)|f_I(q,p)=\mu_I; I=1,\dots,n-k
           \right\rbrace ,\eeqn
      with $\underline{\mu}:=(\mu_1,\dots,\mu_{n-k})$,
      are integral manifolds of $V_{\rV }$. Here, $\rV $ denotes
      the degree of non-holonomy of $V$, and evidently,
      $\Mm_{\underline{0}}=\phys$.

      Condition (iii) is satisfied because
      $$L(q,p)\;:=\;\sum_I\;\left|f_I(q,p)\right|^2$$
      is an integral of motion for orbits of $\tPhi_t$.
      Since $L$ grows monotonically with increasing $|\underline{\mu}|$,
      and attains its (degenerate) minimum of value zero on
      $\phys$, it is a Lyapunov function for $\phys$.
      Anything better than
      marginal stability is prohibited by energy conservation.
\item To satisfy condition (ii), we demand that  $\bbeta(q)\dot{q}=0$,
      or equivalently, that
      \eqn\label{nonhcon}\zeta_I(\dot{q})\;=\;0\;\;\;,\;\;
           \forall I=1,\dots,n-k\;,\eeqn
      shall be satisfied
      along all orbits $(q(t),p(t))$ of (~\ref{eqsofmo}),
      owing to (~\ref{qdotcon}).
\item If the constrained mechanical system exhibits a $G$-symmetry,
      characterized by a group action $\psi:G\rightarrow {\rm Diff}(Q)$
      so that $\psi_{h*}W=W\;\forall h\in G$, the local family of 1-forms
      $\lbrace\zeta_I\rbrace$ can be picked in a manner that
      $\psi_{h}^*\zeta_I=\zeta_I$ is satisfied for all
      $h\in G$ in a vicinity of the unit element $e$.
      Consequently, the functions
      $f_I(q,p)=h^*_q(\zeta_I,p)$ and their level sets
      $\Mm_{\underline{\mu}}$ are invariant under the group action.
\end{enumerate}

The condition that (~\ref{invman})
are integral manifolds of $V_{\rV }\supset V$ implies
that all sections of $V$ are
annihilated by the 1-forms $df_I$, for $I=1,\dots,n-k$.
Furthermore, the condition (~\ref{nonhcon}) requires $V$ to
be annihilated by the 1-forms
\eqn\label{defxi}\xi_I\;:=\;\zeta_{Ir}(q)\;dq^r\;+\;\sum_s\;0\;dp_s\;\eeqn
that are
obtained from lifting $\zeta_I$ to $T^*(T^*Q)$, with $I=1,\dots,n-k$.

\begin{prop}\label{defV}
The distribution
$$V\;:=\;\Big(\bigcap_I{\rm ker}\;df_I\Big)\;\;\;\bigcap\;\;\;
      \Big(\bigcap_I{\rm ker}\;\xi_I\Big)\;\;\subset\;T(T^*Q)$$
is symplectic.
\end{prop}

\begin{proof}
$V$ is symplectic iff its symplectic
complement $V^\perp$ is.
With the given data, the latter condition is more convenient to check.
$V^\perp$ is locally spanned by the vector fields
$(Y_1,\dots,Y_{2k})$ obtained from
\eqn\label{granite}\omega_0(Y_I,\cdot)\;=\;\xi_I(\cdot)\;\;\;,\;\;\;
      \omega(Y_{I+k},\cdot)\;=\;d f_I(\cdot) ,\eeqn
where $I=1,\dots, k$, and $\omega_0=-dp_i\wedge dq^i$.

$V^\perp$ is symplectic if and only if
$D:=\;[\omega(Y_I,Y_J)]$ has values in $GL_\R(2(n-k))$.

We remark that in the present notation, capital indices range from
$1$ to $k$ if they label 1-forms, and from $1$ to
$2k$ if they label vector fields.

In local bundle coordinates,
$$\label{kabana}df_I \;=\; (\partial_{q^i}f_I)(q,p)\; dq^i \;\;+\;\;
      \zeta_{Ii}(q)\;g^{ij}(q)\;dp_j \;,$$
where $g_{ij}$ are the components of the metric tensor $g$ on $Q$,
as before.
Let us introduce the functions
$E(q):=[\zeta_{Ji}(q)]$ and $F(q,p):=[\partial_{q^j}f_K(q,p)]$, both
with values in ${\rm Mat}_\R(n\times (n-k))$,
which we use to  assemble
\eqnn K\;:=\;\left(\begin{array}{cc}
         E^\dagger&0\\F^\dagger&E^\dagger g^{-1}\end{array}\right)\;\;\;\;:\;
         T^*Q\;\longrightarrow\;{\rm Mat}_\R(2(n-k)\times 2n)\;.\eeqnn
Any component vector
$v:T^*Q\rightarrow \R^{2n}$ that locally represents an element
of $\Gamma(V)$ satisfies $K v=0$.
The symplectic structure $\omega_0$ is locally represented by
$J$, defined in (~\ref{sympJ}).
One can easily verify that the $I$-th row vector of the matrix
$K \Jj^{-1}$ is the component vector of $Y_I$.
In conclusion, introducing the matrices
\eqnn G(q)&:=&E^\dagger(q)\; g^{-1}(q)\;E(q)\\
      S(q,p)&:=& F^\dagger(q,p)\;
     g^{-1}(q)\;E(q)\;-\; E^\dagger(q)\;
     g^{-1}(q)\;F(q,p)\;  ,\eeqnn
one immediately arrives at
\eqn D& =&K\Jj K^\dagger\;=\;
     \left(\begin{array}{cc} 0& G\\-G&S
     \label{Dmat}\end{array}\right).\eeqn
Since $\zeta_I$ has been picked a $g^*$-orthonormal family
of 1-forms on $Q$, it is clear that $G(q)=\1_{n-k}$.
Thus, $D$ is invertible. This proves that
$V^\perp$ is symplectic. \end{proof}

\subsubsection{ Construction of the projection tensors}

Next, we determine the matrix  of the
$\omega_0$-orthogonal projection tensor $\PV$, which is associated to $V$, in
the present bundle chart.
Again, it is more convenient
to carry out the construction for its complement first.

\begin{prop}
The matrix of the $\omega_0$-orthogonal projection tensor $\PVc$
associated to $V^\perp$ (considered as a tensor field that
maps $\Gamma(T(T^*Q))$ to itself, with kernel $V$) is given by
\eqnn \PVc=\left(\begin{array}{cc}\bbeta&0\\
      T&\bbeta^\dagger\end{array}\right)\eeqnn
in the local bundle chart $(q,p)$. The matrix $T=T(q,p)$
is defined in (~\ref{eqncs}).
\end{prop}

\begin{proof}
The proof of lemma {~\ref{Pconstr}}
can be used for this proof.
The inverse of (~\ref{Dmat}) is
\eqnn D^{-1}=\left(\begin{array}{cc} S
             &-\1_{n-k}\\
             \1_{n-k}&0 \end{array}\right)\;\;,\eeqnn
where we recall that $G(q)=\1_{n-k}$.
The $I$-th column vector of the matrix
$K \Jj^{-1}$ is the component vector of $Y_I$ (we have required that
$\lbrace Y_1,\dots Y_{2(n-k)}\rbrace$ spans $V^\perp$).
This implies that $\PVc=\Jj K^\dagger D^{-1} K$.

\begin{lem}
The matrix of  $\bbeta$   in the given chart is given by
\eqn\label{betaeq}\bbeta(q)\;=\;g^{-1}(q)\;E(q)\;E^\dagger(q)\;.\eeqn
\end{lem}

\begin{proof}
The construction presently carried out
for $\PVc$ can also be applied to
$\bbeta$. One simply replaces $V^\perp$ by
$W^\perp$, and $\omega_0$ by the Riemannian
metric $g$ on $Q$.
An easy calculation immediately produces the asserted
formula.
The matrix of $\alph$ is subsequently obtained from $\alph+\bbeta=\1$.
For more details, cf. \cite{Bra}. \end{proof}

Introducing
\eqn\label{eqncs} T(q,p)\;:=\;E(q)\;F^\dagger(q,p)\;\alph(q)\;\;-\;\;
     \alph^\dagger(q)\;F(q,p)\; E^\dagger(q)\; ,\eeqn
a straightforward calculation produces the asserted formula
for $\PVc$. \end{proof}

\begin{cor}
In the given bundle coordinates, the matrix of $\PV$ is
\eqnn \PV=\left(\begin{array}{cc}\alph&0\\-T&\alph^\dagger
      \end{array}\right)\;, \eeqnn
where $T=T(q,p)$ is defined in (~\ref{eqncs}).
\end{cor}

\begin{proof}
This is obtained from
$\PV+\PVc =\1_{2n}$. \end{proof}

In this chart, $\PV(x)\Jj=\Jj\PV^\dagger(x)$, by
$\omega_0$-skew orthogonality of $\PV$.

\begin{thm}
Let $H$ be as in (~\ref{hamham}).
Then, the dynamical system locally represented by
\eqn\label{salvation}\left(\begin{array}{c}\dot{q}\\\dot{p}\end{array}\right)
    =\left(\begin{array}{cc}0&\alph\\-\alph^\dagger&-T\end{array}\right)
       \left(\begin{array}{c}\partial_q H\\
       \partial_p H\end{array}\right) \;, 
\eeqn
corresponding to the contrained Hamiltonian system $(T^*Q,\omega_0,H,V)$,
is an extension of the constrained mechanical system $(Q,g,U,W)$.
\end{thm}

\begin{proof}
By construction, $\phys$ is an invariant manifold of the
associated flow $\tPhi_t$, hence
(~\ref{physleaf}) is fulfilled for all orbits of (~\ref{salvation})
with initial conditions in $\phys$.

The equation
$\dot{q}=\alph \partial_p H$ in (~\ref{salvation})
obviously is (~\ref{qdotcon}).

Next, using the notation
$\underline{f}:=\left(f_1,\dots,f_{n-k}\right)^\dagger$,
\eqnn\underline{f}\;=\;E^\dagger\; g^{-1}\;p\;,\eeqnn
and substituting (~\ref{eqncs}) for $T(q,p)$, the equation
for $\dot{p}$ in (~\ref{salvation}) becomes
\eqnn\dot{p}\;=\;-\alph^\dagger\; \partial_q H\;-\;E\; F^\dagger\dot{q}
     \;+\;\alph^\dagger F \underline{f}\;.\eeqnn

Since $M_{\underline{\mu}}$ are invariant manifolds
of the flow $\tPhi_t$ generated by (~\ref{salvation}),
$\partial_t f_I(q(t),p(t))$ vanishes along all orbits of (~\ref{salvation}),
so that
$F^\dagger\dot{q}+E^\dagger g^{-1}\dot{p}=0$.
This implies that
\eqn\dot{p}\;=\;-\alph^\dagger\;\partial_q H\;+\;
       E\; E^\dagger\; g^{-1}\; \dot{p}
      \;+\;\alph^\dagger\;\partial_q \Big(\frac{1}{2}\underline{f}^\dagger
       \underline{f}\Big)\;.\label{dotpeq1}\eeqn
Recalling that $\bbeta=g^{-1}E E^\dagger$ from
(~\ref{betaeq}), and using the fact that $\underline{f}=
\underline{0}$ on $\phys$, one  arrives at
(~\ref{pdotcon}) by multiplication with $\alph^\dagger$ from the left.
\end{proof}

\subsubsection{Equilibria of the extension} 

The critical set of the extension
constructed above is characterized by the following theorem.

\begin{thm}\label{critsetextthm}
The critical set of (~\ref{salvation}) is given by the vector bundle
$$\crit\;=\;\bigcup_{q\in \crit_Q}\{q\}\times (W_q^*)^\perp
      \; \;\; $$
with base space $\crit_Q$, cf. (~\ref{pegasus}).
\end{thm}

\begin{proof}
Let us first consider (~\ref{dotpeq1}).
As has been stated above, the second term on its right hand equals
$\bbeta^\dagger(q)\dot{p}$, and moreover, from (~\ref{holcon}), one concludes that
$$\underline{f}^\dagger
    \underline{f}\;=\; \|\bbeta^\dagger p \|_{g^*}^2\;.$$
The Hamiltonian (~\ref{hamham}) can be decomposed into
$$H(q,p)\;=\;H(q,\alph^\dagger p)\;+\;
    \frac{1}{2}\|\bbeta^\dagger p\|_{g^*}^2\;,$$
due to the $g^*$-orthogonality of $\alph^\dagger$ and $\bbeta^\dagger$,
so that (~\ref{dotpeq1}) can be written as
$$\dot{p}\;=\;-\;\alph^\dagger \partial_q H(q,\alph^\dagger p)\;+\;\bbeta^\dagger\dot{p}\;.$$

The equilibria of (~\ref{salvation}) are
therefore determined by the conditions
$$\alph^\dagger(q)p\;=\;0\;\;\;,\;\;\;\alph^\dagger(q)\;
      \partial_q H(q,\alph^\dagger p)\;=\;0\; .$$
Because $H$ depends quadratically on $\alph^\dagger p$,
the second condition
can be reduced to
$$\alph^\dagger(q)\;\partial_q U(q)\;=\;0$$
using the first condition.
Comparing this with (~\ref{pegasus}), the assertion follows.
\end{proof}

In particular, this fact implies that every
equilibrium $(q_0,p_0)$ of the extension defines a unique
equilibrium $q_0$ on $\crit_Q$ by
base point projection.

To analyze the stability of a given equilibrium solution $q_0\in \crit_Q$,
it is necessary to
determine the spectrum of the linearization of $X_H^V$ at $\xa=(q_0,0)$.

A straightforward calculation along the lines of the previous discussion
shows that in the present bundle chart,
\eqn\label{DXHVlinexpl}DX_H^V(\xa)\;=\;\left(\begin{array}{cc}0&\alph
    g^{-1}\alph^\dagger\\-\alph^\dagger D^2_{q_0}
     U\alph-R&0
   \end{array}\right)(\xa)\; ,
\eeqn
where
\eqn\label{corrosion}\left[R_{jk}\right]\;:=\;
      \left[\;\partial_{q^i}U(\alph)^r_j\;(\alph)^s_k\;
      \partial_{q^s}(\alph)^i_r\;\right]\;\;\;\;
      \in\;{\rm Mat}_\R(n\times n)\; .\eeqn
Furthermore, $D^2_{q_0} U$ is the matrix of second derivatives of $U$.
The stability discussion in the previous
section can now straightforwardly be applied to
$DX_H^V(\xa)$.

\subsubsection{ Extension of symmetries}

Let us assume that  the constrained mechanical system $(Q,g,U,W)$
exhibits a $G$-symmetry $\psi:G\rightarrow {\rm Diff}(Q)$.
Then, we claim that it is extended by $(T^*Q,\omega_0,H,V)$. 
To this end, we recall that the 1-forms $\zeta_I$ satisfy
$\psi^*_h \zeta_I$ for all $h\in G$ close to the unit.

Via its pullback, $\psi$ induces the group action
\eqnn\Psi:=\psi^*\;\;:\;\;G\times T^*Q&\longrightarrow& T^*Q\eeqnn
on  $T^*Q$.
This group action is symplectic, that is,  $\Psi_h^*\omega_0=\omega_0$
for all $h\in G$.
For a proof, consider for instance \cite{AbMa}.

The 1-forms $\xi_I$, defined in (~\ref{defxi}), satisfy
$\Psi_h^*\xi_I=\xi_I$, and likewise,
$f_I\circ\Psi_h=f_I$ is satisfied for all $h\in G$ close to the unit.
The definition
of $V$ in proposition {~\ref{defV}} thus implies that
$$\Psi_{h\,*}V\;=\;V$$
is satisfied for all $h\in G$.
Due to the fact that $\omega$ and $V$ are both $G$-invariant,
$\PV$ and $\PVc$ are also invariant under the $G$-action $\Psi$.

The Hamiltonian $H$ in (~\ref{hamham}) is $G$-invariant under $\Psi$,
by assumption on the constrained Hamiltonian
mechanical system.
Thus,  $X_H$ fulfills
$\Psi_{h*}X_H=X_H$ for all $h\in G$, which implies that
$X_H^V=\PV (X_H)$ is $G$-invariant.

\subsection{\nbf The Topology of the Critical Manifold}

Since
$\crit$ is   not  a compact submanifold of $T^*Q$, our
previous results cannot be applied directly. However, owing to
the vector bundle structure of $\crit$ and $T^*Q$, the result
\eqn\label{eqnhd} \sum_{i,p} \lambda^{p+\mu_i} {\rm dim}H^p_c(\crit_i) =
     \sum_{p} \lambda^p {\rm dim}H^p_c(T^*Q) +(1+\lambda) \Q(\lambda) \;
\eeqn
still holds,
where $H^*_c$ denotes the de Rham cohomology based on differential forms with
compact supports.  The polynomial
$\Q(t)$ has non-negative integer coefficients.

In a first step, the arguments of section {~\ref{sectionII}} can be straightforwardly
applied to  $\crit_Q$. $\crit_Q$ is normal hyperbolic
with respect to the gradient-like flow $\psi_t$ generated by
$$\partial_t q(t)=-\alph(q(t))\nabla_g U(q(t)) ,$$
it contains all
critical points of the Morse function $U$, but no other conditional extrema
of $U|_{\crit_Q}$ apart from those (it is gradient-like because along all of
its non-constant orbits, $\frac{d}{dt}U(t)=-g(\alph\nabla_g
U,\alph\nabla_g U)|_{q(t)}<0$ holds, since $\alph$ is an orthogonal 
projection tensor with respect to the Riemannian metric $g$ on 
$Q$).  This can be proved by substituting $M\rightarrow Q$, $H\rightarrow U$,
$\PV\rightarrow\alph$, $g_{({\rm Kahler})}\rightarrow g$, and
$\crit\rightarrow\crit_Q$ in section {~\ref{sectionII}}, and by applying the 
arguments used there.
Hence, letting $\mu_i$ denote the index of the connectivity component
$\crit_{Qi}$ of $\crit_Q$, (~\ref{Conley-Zehnder1}) implies that 
for compact, closed $Q$, 
\eqn\label{eqnhdbum} \sum_{i,p} \lambda^{p+\mu_i} {\rm dim}H^p(\crit_{Qi}) =
     \sum_{p} \lambda^p {\rm dim}H^p(Q) +(1+\lambda) \Q(\lambda) \;, 
\eeqn
where $\Q(t)$ is a polynomial with non-negative integer coefficients.

$\crit_Q$, being the zero section of $\crit$,
is a deformation retract of $\crit$,
and likewise, $Q$ is a deformation retract of $T^*Q$.
Thus, (~\ref{eqnhd}) follows trivially from the 
invariance of the de Rham cohomology groups under retraction,
$H^p_c(\crit_i)\cong H^p(\crit_{Qi})$, $H^p_c(T^*Q)\cong H^p(Q)$.
Hence, (~\ref{eqnhd}) is equivalent to
\eqn\label{MorseBottCZQcase}  \sum_{i,p} \lambda^{p+\mu_i} b_p(\crit_{Qi}) =
      \sum_{p} \lambda^p b_p(Q) +(1+\lambda) \Q(\lambda),\eeqn
where $b_p$ is the $p$-th Betti number.

Consequently, one finds
$\sum_{i} b_{p-\mu_i}(\crit_{Qi}) \geq b_{p}$, and
in particular, for $\lambda=-1$, one obtains
$$\sum_{i,p} (-1)^{p+\mu_i} b_p(\crit_{Qi}) =
         \sum_{i} (-1)^{\mu_i} \chi(\crit_{Qi}) =\chi(Q) ,$$
where $\chi$ denotes the Euler characteristic.

\section{Applications, Illustrations and Examples}

Let us conclude our analysis with the discussion of some simple 
applications and examples.

\subsection{\nit A Computational Application}

Let us first formulate an application of our analysis
for the computational problem of
finding the  equilibria in a large
constrained multibody system.
It is in this context also desirable to determine whether a given
set of parameters and constraints implies the existence of
non-generic critical points. This is due to the circumstance that in practice, 
manufacturing imprecisions can have a significant 
effect on the latter.

For large multibody systems, 
equilibria can realistically only be determined by numerical routines.
The strategy presented in chapters {~\ref{sectionII}} and {~\ref{sectionIV}}
suggests the following method.

If $U$ is a Morse function
whose critical points are known,
and if $Q$ is compact and closed,
it is possible to numerically construct all generic
connectivity components of $\crit_Q$.
This is because generic components of
$\crit_Q$ are smooth, $n-k$-dimensional submanifolds of $Q$ containing all
critical points of $U$, and
no other critical points of $U|_{\crit_Q}$.
This information can be exploited to find sufficiently many points
on $\crit_Q$, so that a suitable interpolation routine enables the
approximate reconstruction of an entire connectivity component.
To this end, one chooses a vicinity of a critical point $a$ of $U$, and
uses a fixed point solver to determine neighboring zeros of
$|\alph(q)\nabla_g U(q)|^2$, which are
elements of $\crit_Q$ close to $a$.
Iterating this procedure with the critical points found in
this manner, pieces of $\crit_Q$ of arbitrary size can be
determined.

If all critical points of $U$ are a priori known,
one can proceed in this manner to construct all connectivity
components of $\crit_Q$ that contain critical points of $U$.
Then,
one is guaranteed to have found all of the generic components of $\crit$ if
the numerically determined connectivity components are
closed, compact, and contain all critical points of $U$.

We remark that determining the critical points of
a Morse function $U:Q\rightarrow\R$
is a difficult numerical
task by itself. Attempting to find critical points by simulating the
gradient flow generated by $-\nabla_g U$ is  
time costly, because the critical points define a
thin set in $M$.
Their existence, however, is of course ensured by the topology of $Q$.

Another remark is that all critical points $a$
at which $D(\alph \nabla_g U)(a)$
has a reduced rank, are elements of the non-generic
part of $\crit_Q$. Thus, the latter condition is an indicator
for non-genericity.
If there are such exceptional critical points in a technically relevant
region of $Q$, they can be removed by a small
local modification of the system parameters or constraints.

\subsection{A disc in a periodic potential, sliding on the plane}

Let us consider a mechanical example,
consisting of a thin disk of radius $r$ and mass $m$ on the plane
$\R^2$, which is attached
to a massless skate. The connecting line between the center of the disc
and the contact point at the center
of the skate with the plane is normal to the plane, precisely
if the disc is horizontal. 
We assume that the disk remains horizontal during its motion,
and that the translational
motion of the disc is only possible in the direction of the skate.

Let $(x_1,x_2)$ denote the position of the center of mass of the 
disc with respect to some Euclidean coordinate system on $\R^2$,
and let $\phi$ denote the angle enclosed by the skate and the $x_1$-axis.

The kinetic energy of this system is given by
$$
		T=\frac{m }{2}(\dot{x}_1^2+\dot{x}_2^2)
		+\frac{1}{2}\frac{mr^2}{2}
       \dot{\phi}^2 \; ,
$$
which defines a Riemannian metric on $TQ$ with metric tensor
$$
		[g_{ij}(\phi,\theta,\psi)]=
		\left(\begin{array}{ccc}m &0&0\\
		0&m &0
		\\0&0&\frac{mr^2}{2}
   	\end{array}\right) . $$
Furthermore, we assume that it moves against the background of a 
$(2\pi\Z)^3$-periodic potential energy  
$$U(x_1,x_2,\phi)=\sum_{i=1,2} c_i (1-\cos x_i)+  c_\phi (1-\cos\phi) \; ,$$
where $c_1$, $c_2$, and $c_\phi$ are coupling constants.  
 
Dividing out the translational symmetry with respect to $(2\pi\Z)^3$,
the configuration manifold of this mechanical system is given
by $Q=[0,2\pi]^3\cong  T^3$
(periodic boundary conditions). Clearly, $U:T^3\rightarrow \R$ is a real analytic
Morse function, with 8 critical points in the corners of $[0,\pi]^3$,
while each of the remaining critical points in $[0,2\pi]^3$ is identified
with one of the former by periodicity. Correspondingly, we will from
here on consider $(x_1,x_2)$ as coordinates on $T^2$, that is, mod $2\pi$.

The requirement that the disk shall slide in the direction of the
skate is expressed by the non-holonomic constraint  
$$\dot{x}_1\sin\phi- \dot{x}_2\cos\phi=0 .$$
The matrix $E^\dagger(x_1,x_2,\phi)$, introduced in the 
proof of theorem {~\ref{defV}}, thus corresponds to 
$$E^\dagger(x_1,x_2,\phi)=(\sin\phi,-\cos\phi,0)\;,$$ 
so that  $E^\dagger g^{-1}E=\frac{1}{m}$. 

The orthoprojectors $\bbeta$ and $\alph$ are thus straightforwardly 
obtained as
\eqnn
	\bbeta(x_1,x_2,\phi)&=&\left(\begin{array}{ccc}
	\sin^2\phi&-\sin\phi\cos\phi&0
	\\ -\sin\phi\cos\phi&\cos^2\phi&0\\0&0&0
   \end{array}\right) \\
   \alph(x_1,x_2,\phi)&=&\left(\begin{array}{ccc}
 	\cos^2\phi&\sin\phi\cos\phi&0
	\\ \sin\phi\cos\phi&\sin^2\phi&0\\0&0&1
   \end{array}\right) .\eeqnn
The critical set is given by
$$
		\crit_Q=\Big\{(x_1,x_2,\phi)\Big|\big(\alph^\dagger \nabla U\big)(x_1,x_2,\phi)
		=0\Big\}
$$ 
(where $\nabla:=(\partial_{x_1},\partial_{x_2},\partial_\phi)$). 
Let 
$$
		\crit_{a,b}:=\Big\{(x_1,x_2,\phi)\Big|x_1=a \, , \,
		x_2\in[0,2\pi]\, , \, \phi=b\Big\}\;.
$$
Then,  
$$
		\crit_Q=\bigcup_{a,b\in\{0,\pi\}}\crit_{a,b} \;.
$$ 
It is trivially clear that $\crit_Q$ contains all critical points of $U$.
Let $q_c\in\crit_{a,b}$, where $a,b\in\{0,\pi\}$. 
Noting that $\alph={\rm diag}(1,0,1)$ on $\crit_Q$, we have
\eqn\label{icestorm}
	\Big(\nabla\otimes(\alph^\dagger \nabla U)\Big)(q_c)
	&=&\Big(\alph^\dagger \big( \nabla\otimes\nabla U\big)\alph\Big) 
	(q_c)+\tilde{R}(q_c)
	\nonumber\\
	&=&
   \left(\begin{array}{ccc}
  	c_1\cos a&0&c_2\sin x_2
  	\\ 0&0&0\\0&0&c_\phi \cos b
	\end{array}\right) \;. \eeqn
Clearly, 
$$
	{\rm spec}\Big(\Big(\nabla\otimes(\alph^\dagger \nabla U)\Big)(q_c)\Big)
	=\Big\{0,c_1\cos a,c_\phi \cos b\Big\} \;,
$$
which is, for each fixed $a,b$, independent of $x_2$. Thus, the indices
of the connectivity components $\crit_{a,b}$ with respect to the gradient-like
flow generated by $-\alph^\dagger \nabla U$ are given by
\eqn
		\mu(\crit_{0,0})=2\; \; , \; \; \mu(\crit_{0,\pi})=\mu(\crit_{\pi,0})=1
		\; \; , \; \; \mu(\crit_{\pi,\pi})=0 \;,
\eeqn
and clearly, $\crit_{a,b}\cong S^1$ for all $a,b\in\{0,\pi\}$. Since the
Betti numbers of $T^3$ are given by $b_0=b_3=1$, $b_1=b_2=3$, 
and those of $\crit_{a,b}$ by $b_0(\crit_{a,b})=b_1(\crit_{a,b})=1$,
$b_2(\crit_{a,b})=b_3(\crit_{a,b})=0$, one
finds that 
$$
		\sum_{a,b} b_{p-\mu(\crit_{a,b})}(\crit_{a,b})=b_p(Q)
$$
for $p=0,\dots,3$, or explicitly,
\eqnn
		b_{3-2}(\crit_{0,0})&=&1=b_3(T^3)\\
		b_{2-2}(\crit_{0,0})+b_{2-1}(\crit_{0,\pi})+b_{2-1}(\crit_{\pi,0})&=&3
		=b_2(T^3)\\
		b_{1-0}(\crit_{\pi,\pi})+b_{1-1}(\crit_{0,\pi})+b_{1-1}(\crit_{\pi,0})
		&=&3=b_1(T^3)\\
		b_{0-0}(\crit_{\pi,\pi})&=&1=b_0(T^3)\;,
\eeqnn
in agreement with (~\ref{MorseBottCZQcase}).
	
Next, we determine the spectrum of the linearization of $X_H^V$
at $(q_c,0)\in T^*Q$, cf. (~\ref{DXHVlinexpl}).
To this end,
$$
	\Big(\alphc g^{-1}\alphc^\dagger\Big)(q_c)
	 =\left(\begin{array}{ccc}
    \frac1m&0&0\\0&0&0
    \\0&0&\frac{2}{mr^2}
   \end{array}\right) ,$$
and multiplying this matrix from
the right with (~\ref{icestorm}) yields
$$
	\Omega(q_c,\theta):=\left(\begin{array}{ccc}
    \frac{c_1\cos a}{m}&0&
	\frac{c_2 \sin x_2}{m}
    \\0&0&0\\0&0& \frac{2 c_\phi \cos b}{mr^2} 
   \end{array}\right) .$$
Clearly,
$$
		{\rm spec}\Big(\Omega(q_c,\theta)\Big) = 
		\Big\{0,\frac{c_1\cos a}{m},\frac{2 c_\phi \cos b}{mr^2}\Big\} \;.
$$
From (~\ref{DXHVlinexpl}), it is easy to see that
$$
		{\rm spec}\Big(DX_H^V(q_c,0)\Big)=
		\Big\{0,\pm\sqrt{\frac{c_1\cos a}{m}},
		\pm\sqrt{\frac{2 c_\phi \cos b}{mr^2}}\Big\}\;,
$$
hence critical stability occurs for the case $a=b=\pi$, while
in all other cases, there is an asymptotically unstable direction.

We conclude that all components $\crit_{a,b}$, where $a+b\leq \pi$,  are unstable.
In the critically stable case $a=b=\pi$, the linear problem is oscillatory,
and the eigenfrequencies  are given by $\sqrt{\frac{c_1}{m}}$ and 
$\sqrt{\frac{2 c_\phi}{mr^2}}$, independently of $x_2$. Since $\mu(\crit_{\pi,\pi})=0$, 
our discussion 
in section {~\ref{sectionIII}} suggests that the connectivity component $\crit_{\pi,\pi}$
of $\crit_Q$ is stable in the sense of Nekhoroshev
if the ratio $\sqrt{\frac{c_1 mr^2}{2c_\phi}}$
is irrational.

\subsection*{\bf Acknowledgements}

This work is based on the thesis \cite{Ch}, 
which was carried out at the center
of mechanics (IMES), ETH Z\"urich.
I warmly thank  Prof. H. Brauchli for
suggesting this area of problems, for his insights, 
and for the possibility to 
carry out this work.
I am profoundly grateful to Prof. E. Zehnder for his generosity,
and discussions that were most enlightening
and helpful.
It is a pleasure to thank
M. von Wattenwyl, M. Sofer, H.
Yoshimura,    
O. O'Reilly, and especially  M. Clerici, for highly interesting discussions.
I also thank the referee for his helpful suggestions.
The author is supported by a Courant Instructorship.

\end{document}